\documentclass[useAMS,usenatbib]{mn2e}

\usepackage[dvips]{graphicx}   
\usepackage{times}

\usepackage{epsfig}
\usepackage{latexsym}
\usepackage{graphicx}
\usepackage{dcolumn}
\usepackage{bm}
\usepackage{epstopdf}
\usepackage{natbib}
\def\reff@jnl#1{{\rm#1\/}}
\def\aj{\reff@jnl{AJ}}                  
\def\araa{\reff@jnl{ARA\&A}}            
\def\apj{\reff@jnl{ApJ}}                
\def\apjl{\reff@jnl{ApJ}}               
\def\apjs{\reff@jnl{ApJS}}              
\def\ao{\reff@jnl{Appl.Optics}}         
\def\apss{\reff@jnl{Ap\&SS}}            
\def\aap{\reff@jnl{A\&A}}               
\def\aapr{\reff@jnl{A\&A~Rev.}}         
\def\aaps{\reff@jnl{A\&AS}}             
\def\azh{\reff@jnl{AZh}}                
\def\baas{\reff@jnl{BAAS}}              
\def\jrasc{\reff@jnl{JRASC}}            
\def\memras{\reff@jnl{MmRAS}}           
\def\mnras{\reff@jnl{MNRAS}}            
\def\pra{\reff@jnl{Phys.Rev.A}}         
\def\prb{\reff@jnl{Phys.Rev.B}}         
\def\prc{\reff@jnl{Phys.Rev.C}}         
\def\prd{\reff@jnl{Phys.Rev.D}}         
\def\prl{\reff@jnl{Phys.Rev.Lett}}      
\def\pasp{\reff@jnl{PASP}}              
\def\pasj{\reff@jnl{PASJ}}              
\def\qjras{\reff@jnl{QJRAS}}            
\def\skytel{\reff@jnl{S\&T}}            
\def\solphys{\reff@jnl{Solar~Phys.}}    
\def\sovast{\reff@jnl{Soviet~Ast.}}     
\def\ssr{\reff@jnl{Space~Sci.Rev.}}     
\def\zap{\reff@jnl{ZAp}}                
\def\nat{\reff@jnl{Nature}}             
\newcommand{\be}{\begin{equation}}
\newcommand{\ee}{\end{equation}}
\newcommand{\bea}{\begin{eqnarray}}
\newcommand{\eea}{\end{eqnarray}}
\newcommand{\bi}{\begin{itemize}}
\newcommand{\ei}{\end{itemize}}
\title%
[The Faint Source Correlation Function]
{A Measurement of The Faint Source Correlation Function in the GOODS and UDF Surveys}
\label{firstpage}

\author[Morganson et al.]
{Eric Morganson$^{1}$ and
Roger Blandford$^{1}$\\
$^{1}$KIPAC, P.O. Box 20450, MS29, Stanford, CA 94309, USA}

\date{Accepted ---; received ---; in original form \today}

\pagerange{\pageref{firstpage}--\pageref{lastpage}}

\pubyear{2008}

\begin{document}

\maketitle


\begin{abstract}
We present a stable procedure for defining and measuring the two point angular
autocorrelation function, $w(\theta) = (\theta/\theta_0(V))^{-\Gamma}$, of faint ($25 < V
< 29$), barely resolved and unresolved sources in the HST GOODS and UDF datasets. We
construct catalogs that include close pairs and faint detections. We show, for the first
time, that, on subarcsecond scales, the correlation function exceeds unity.  This
correlation function is well fit by a power law with index $\Gamma \approx2.5$ and a
$\theta_0 = 10^{-0.1(V-25.8)}\ \rm{arcsec}$. This is very different from the values of
$\Gamma \approx 0.7$ and $\theta_0(r) = 10^{-0.4(r-21.5)}\ \rm{arcsec}$   associated
with the gravitational clustering of brighter galaxies. This observed clustering probably
reflects the presence of giant star-forming regions within galactic-scale potential
wells. Its measurement enables a new approach to measuring the redshift distribution of
the faintest sources in the sky.
\end{abstract}





\section{Introduction}

The two-point angular and spatial autocorrelation functions, $w(\theta)$ and $\xi(r)$ are
defined as:
\begin{equation}
\left<\frac{dP}{d\vec{\theta}}(\vec{\theta_0})\frac{dP}{d\vec{\theta}}(\vec{\theta_0}+\vec{\theta})\right>=
{\sigma}_0^{2}(1+w(\theta))
\end{equation}
and
\begin{equation}
\left<\frac{dP}{d\vec{r}}(\vec{r_0})\frac{dP}{d\vec{r}}(\vec{r_0}+\vec{r})\right> =
{\rho}_0^{2} (1+\xi(r))
\end{equation}

\begin{figure}
\begin{center}
\centering\includegraphics[width=\linewidth]{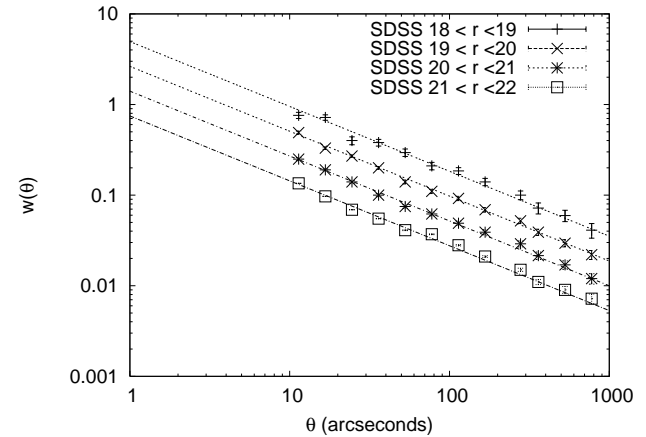}
\caption{The SDSS correlation function with different limiting
magnitudes.}\label{fig:bigW}
\end{center}
\end{figure}

The correlation function of galaxies has been studied observationally at least as far
back as \citet{TOTS1969}. Most recently, the large scale angular correlation function,
$w(\theta)$, function was measured for the galaxies in the Sloan Digital Sky Survey
(SDSS) \citep{CONN2002} for sources with magnitude $18 < r < 22$ and on scales $10'' <
\theta < 1000''$. This and similar work with the 2 Degree Galaxy Redshift Survey (2DF)
\citep{HAWK2003} confirmed that $w(\theta)$ is consistent with a power law form (see Fig.
\ref{fig:bigW}):
\begin{equation}
w(\theta,r) = \left(\frac{\theta}{\theta_0(r)}\right)^{-\Gamma}
\end{equation}
where:
\begin{equation}\label{eq:theta0}
\theta_0(r) = 10^{-0.4(r-21.5)}\ \rm{arcsec};\ \Gamma = 0.72
\end{equation}

$\theta_0$ is roughly proportional to the flux of the source. If we were to extrapolate
this trend out to $r < 25$, we would expect a $\theta_0 = 0.054''$, too small to observe
with current data.

Both the SDSS and 2DF groups measured the spatial correlation function \citep{ZEHA2002}
\citep{HAWK2003}. SDSS observed $r < 22.5$, $\ 0.02 < z < 0.13$ sources on scales of
$0.14$ Mpc $ < r < 23$ Mpc. These observations essentially confirm a power law model of
the correlation function with exponent $\gamma \approx \Gamma +1$, consistent with
\citet{LIMB1953}:

\begin{equation}
\xi(r) = \left(\frac{r}{r_0}\right)^{-\gamma};\ r_0 \approx r_c
\theta_0^{\Gamma/(\Gamma+1)}
\end{equation}
where $r_c$ is the characteristic distance to the sources and the sources are distributed
with a width $\Delta r_c \approx r_c$.

\citet{MASJ2006} extended the SDSS measurement of $\xi(r)$ down to $14$ kpc with moderate resolution difficulties on the smallest scales and found that $\xi(r)$ is consistent with a $\Gamma = 2$ power law over four orders of magnitude.

\citet{PEEB1974} proposed that the cosmological scale correlation function and its power
spectrum counterpart should be related to the microwave background fluctuation using the
theory of gravitational perturbations. On cosmological scales, the dark matter correlation function evolved from
primordial mass fluctuations. Halo Occupation Distribution (HOD) frameworks are used to predict galaxy bias within large dark matter halos \citep{PEAC2000}. Today, theory and experiment are in excellent agreement on the largest scales e.g. \citet{TEGM2004}, and measurements down to $0.3$ Mpc, including slight perturbations from a power law, can be explained in the HOD framework \citep{ZEHA2004}. The continuation of a power law down to smaller scales is less understood, but \citet{MASJ2006} note that it could be accommodated by HOD models with reasonable modifications.

Measurements of $\xi(r)$ are fundamentally restricted to bright sources by the need for
redshifts. HST and large telescopes make spectroscopic redshift measurements good for $r
< 25$ and photometric measurements for $r < 27$ \citep{COE2006}. But the faintest
photometric redshifts cannot be calibrated. Ellipticity measurements are very uncertain
for sources which are not significantly larger than the PSF, which hinders HST
ellipticity measurements for sources dimmer than $r \approx 26$. But even for the
faintest sources, we can precisely measure position and, with a large enough sample,
$w(\theta)$.


The $25 < V < 29$ sources we study in the paper are only $0.1''-0.5''$ in size, no more 
than a few kpc across at any redshift and much smaller than local galaxies. We find that these 
sources are only significantly clustered on subarcsecond scales. Even at high redshifts, the 
physical correlation scales would be roughly $5$ kpc and smaller, much smaller than the scales 
probed by \citet{MASJ2006}. These sources tend to be bluer than the luminous red galaxies (LRG)s selected by the SDSS 
groups, as they were selected in the $V$ band. Blue sources with sub-galactic luminosities and 
sizes separated by sub-galactic distances are likely to be separate star-forming regions in the 
same dark matter potential wells. The cosmological effect which causes the correlation function 
observed in SDSS might have some influence at such small scales, but effects other than gravity - 
gas dynamics, star formation, supernovae and so on- dominate over any simple halo modeling. In 
addition, the faint source correlation function (FSCF) traces luminosity, but its relationship to 
mass is unclear.

Measuring $w(\theta)$ for faint sources on small scales is an important method of probing
how these primarily non-gravitational effects augment the gravitational correlation
function. With much larger datasets, we could study the transition from gravitational to
non-gravitational domination in the correlation function. Despite the fact that this is
not possible with current data, the FSCF is interesting in its own right as a useful
time-dependent tracer of star formation and galactic structure,

The small, faint sources we study are an important astrophysical mystery. If we
extrapolate the source counts from the Hubble Ultra Deep Field (UDF) to the whole sky, we estimate that there are
$\approx 10^{11}$ sources in the sky, and yet if we extend our local galaxy density out
to the $\approx 10^{12}$ Mpc$^3$ comoving volume within $z \leq 4$, we obtain roughly one
tenth this number. These sources are therefore likely to be the subunits of future
galaxies and studying their redshift distribution would be a useful way to probe galaxy
assembly. Unfortunately, we do not know the distance to these sources. Previous estimates
have ranged from $z < 1$ \citep{BABU1992} to $z \approx 2.5$ \citep{HE2000}. Their
dimness and small size make them difficult to study photometrically or geometrically, but
we can study the way they cluster with some precision.

\begin{figure}
\centering\includegraphics[width=\linewidth]{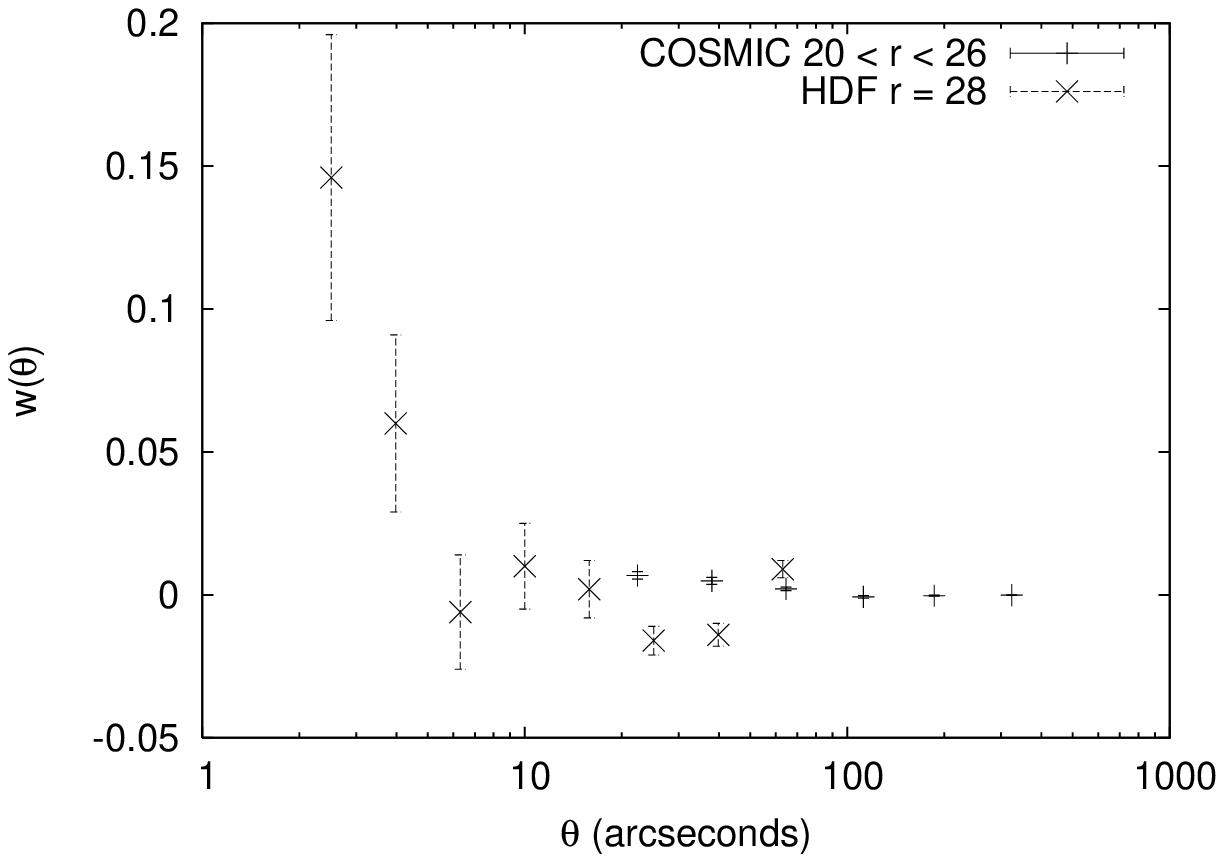}
\caption{The faint correlation function in COSMIC \citep{BRAI1995} and the HDF
\citep{VILL1997}.}\label{fig:oldW}
\end{figure}

In this paper, we observe the FSCF in the HST GOODS and UDF in the $0.3'' < \theta <
10''$ range. The excellent angular resolution allows us to make the first statistically
significant measurement of the $w(\theta)$ for faint sources and to measure both
$\theta_0$ and $\Gamma$. In SDSS, \citet{LI2007} found a significant correlation function
of $r < 17.8$ sources on scales down to $10\ \rm{kpc} \approx 2''$ at their limiting
redshift of $z = 0.3$. \citet{BRAI1995} studied the correlation function of $r < 26$
sources down to scales of $30''$ using the COSMIC imaging spectrograph and showed that they were on the order of $0.01$ at these large scales. \citet{VILL1997}
used $r < 29$ sources on scales down to $3''$ in the Hubble Deep Field (HDF) but did not find a correlation function larger than $0.2$ or more than $2 \sigma$ greater than 0. \citet{CONN1998} used the $i <  27$ sources to to confirm that $w(\theta)$ was on the scale of $0.1$ for arcsecond $\theta \approx 1''$. As
shown in Fig. \ref{fig:oldW}, the groups studying faint sources found 
$w(\theta)$ was equal to only a few tenths and barely statistically
significant.

In section 2 of this paper, we discuss the data used to make these measurements. In
section 3, we explain our computational methods for producing simulated images. In
section 4, we describe the production and characteristics of catalogs used in this
analysis. We discuss how we produce our estimate of the correlation function in section
5. In section 6, we present our best fit models to the data, and in section 7 we discuss
the astrophysical significance of our findings and how they could be applied to future
surveys.

In future papers, we will show that the three-point autocorrelation function is also
measurable for these faint sources. We will provide a formalism to measure how the
two-point correlation function is distorted by a gravitational lens and use it to relate
the distributions of source and lens redshifts to the extent permitted by existing data.
Finally, we will measure this effect and use it to relate the distributions of source and
lens redshifts to the extent permitted by existing data.

\section{Samples}
Previous measurements of the FSCF were limited by low resolution and poor statistics. To
overcome resolution difficulties, we use HST observations with roughly $0.12''$
resolution. To improve upon HDF measurements we use samples which are either larger or
deeper to increase the number of total sources. We measure the positions of the $25 < V <
28$ sources in the HST Great Observatories Origins Deep Survey (GOODS) North and South
\citep{GIAV2004} and make $27 < V < 29$ measurements in the HST UDF \citep{BECK2006}.

The GOODS fields cover roughly 160 arcmin$^2$ each in the ACS \textit{BViz} bands (F435W, F606W,
F814W, and F850LP).The \textit{V} band limiting magnitude for a $10\sigma$ detection of a point
source is 27.8. The standard catalogs made detections in the z band and found 29599
sources in the South and 32048 sources in the North. Using methods detailed in section 4,
we produce catalogs with 56088 and 60182 sources in the South and North respectively.

The UDF covers roughly 11 arcmin$^2$ in the same bands but probes roughly 1.5 m deeper
than GOODS. The \textit{V} band limiting magnitude for a $10\sigma$ detection of a point source is
29.3. A standard catalog was made in the i band with 10,040 sources. Using methods
described in section 4, we produce \textit{V} band catalogs with 7298 sources.
\begin{table}
\begin{tabular}{ccccc}
                & F435W & F606W & F775W & F850LP\\
\hline
GOODS & 27.8 & 27.8 & 27.1 & 26.6        \\
UDF & 29.1 & 29.3 & 29.2 & 28.7        \\

\hline
\end{tabular}
\caption{The limiting magnitudes for $10 \sigma$ detections of point sources in different
bands of GOODS and UDF.}
\end{table}

Studying the angular correlation in HST Cosmic Evolution Survey (COSMOS) would be an interesting extension to this work. COSMOS covers two square degrees and is complete for $0.5$'' sources down to $i = 26$ \citep{SCOV2007}. This is roughly equivalent to a $V = 25$ for typical sources. Measuring the FSCF in COSMOS would allow us to fill in the gap between the $V > 25$ work here and the $r < 22$ work in SDSS. Unfortunately, the enormous size of COSMOS make the simulation techniques we use here impractical, and measuring the COSMOS correlation function will have to be a separate effort.

\section{Producing Simulated Data}

In order to measure the FSCF accurately, we must correct for non-astrophysical
correlation effects like optical resolution limits, the incorrect deblending of sources
of non-zero size and the clustering of noise peaks. Simulated images are our main tool in
estimating these effects and determining how to make the best catalogs for these
observations.

To make simulated data, we generated images with only sources, convolved them with a
simulated HST ACS PSF and added Gaussian noise fields that had been convolved with a
separate noise correlation PSF. We tested and rejected many parameterizations of the
source characteristics, source distribution, noise models and PSFs. This simulation
required that many parameters be fine-tuned to match the statistical properties of GOODS
and UDF. In the following descriptions we describe these parameters, the values we
adopted and the quantitative rationale for these choices.

\subsection{Simulated Source Profiles}\label{subsec:profile}

We compared real ($25 < V < 27$) sources to the best fit de Vaucouleurs, Lorentzian,
exponential and Gaussian profiles. After considerable experimentation, we adopted the de
Vaucouleurs profile, but found that it performed only moderately better than the other
profiles, because the sources are small and barely above threshold. For all fits, the
profile is assumed to be elliptical, and we convert to fit to a circular profile using an
effective radius, $r$ defined by:
\begin{equation}
r^2 = a_1(x-x_0)^2+a_2(y-y_0)^2+a_3(x-x_0)(y-y_0)
\end{equation}
with $(x_0,y_0)$ being fit to define a centre and $(a_1, a_2, a_3)$ being fit to give the
source an arbitrary ellipticity and orientation.

In the case of the de Vaucouleurs profile, we taper the sharp central peak when $r^2 <
0.5 r_{1/2}^2$ where $r_{1/2}$ is the half light radius. In this central region of a few
pixels, we replace $r^2$ with $0.25 (r_{1/2}^2+r^2)$  The sharp peak would be flattened
by the PSF and is in fact not even observed in fully resolved sources \citep{ALAM2002}.

In addition to the above shape parameters, we also fit an integral intensity, $I_C$, and
a simple background with a linearly varying intensity so that the function to which we
fit surface brightness:
\begin{equation}
B(x,y) = b_0+ b_1 x + b_2 y + I_0 I(r^2)
\end{equation}
where I is the normalized intensity of the profile being tested.

\begin{figure*}
\begin{minipage}[t]{0.49\linewidth}
\centering\includegraphics[width=\linewidth]{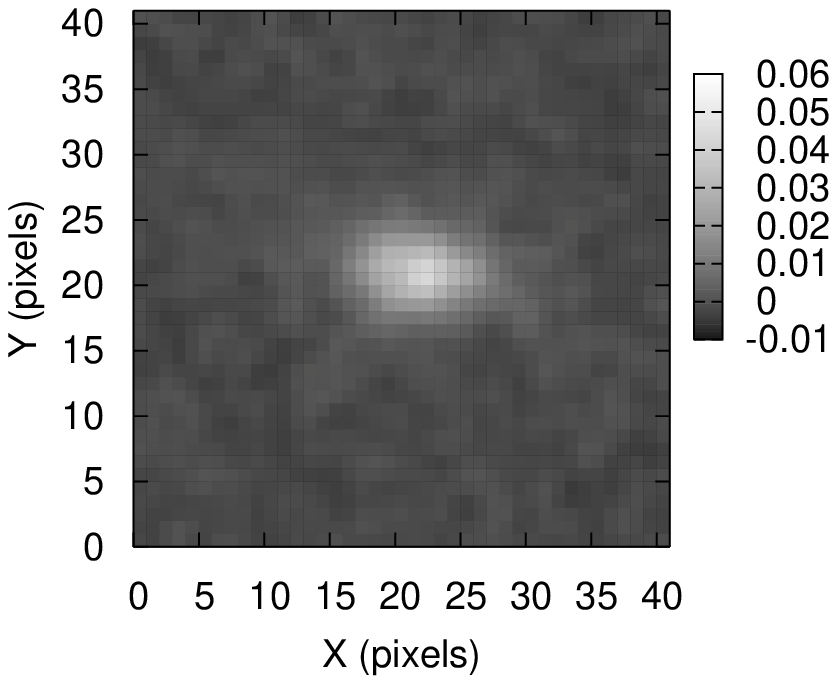}
\end{minipage} \hfill
\begin{minipage}[t]{0.49\linewidth}
\centering\includegraphics[width=\linewidth]{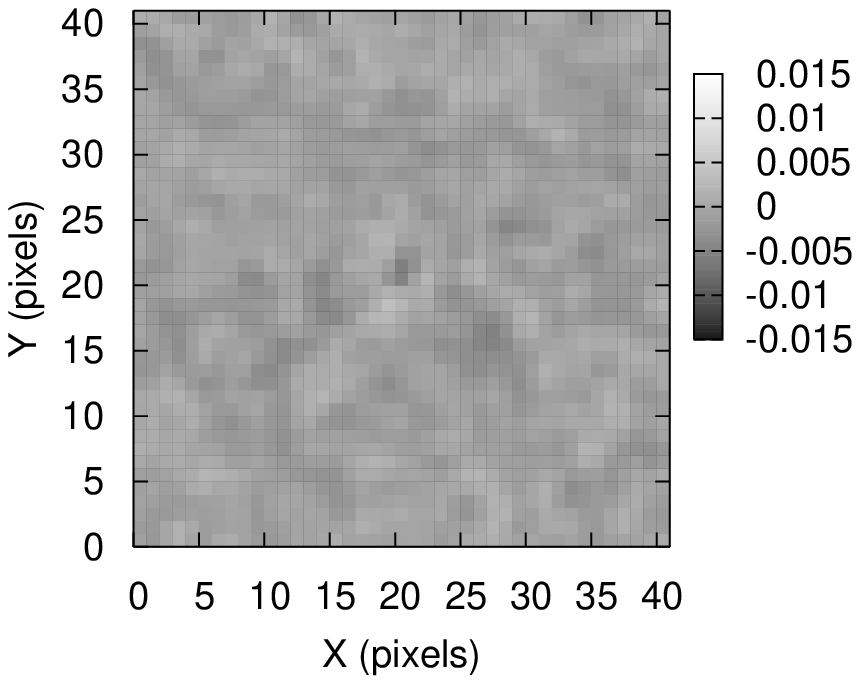}
\end{minipage}
\caption{A typical source profile from real GOODS data (left) and the residuals left by a
de Vaucouleurs fit (right).}\label{fig:DV}
\end{figure*}

For each source the data is fit inside of a square with sides of length $3\sqrt{A}$ where
$A$ is the detection area (a typical source and fit is in Fig. \ref{fig:DV}). We use the
standard deviations implied by the weight images and compare the $\chi^2$ over degrees of
freedom for each fit type. We find that our modified de Vaucouleurs profile has an
average $\chi^2/N_{DoF}$ of 1.386 while the Gaussian, Lorentzian and Exponential profiles
have fit values of 1.396, 1.396 and 1.391 respectively so the choice of profile was not
critical. We use de Vaucouleurs profile but note that the difference in fit quality for
such faint sources is minimal.

\begin{figure}
\begin{minipage}[t]{1.0\linewidth}
\centering\includegraphics[width=\linewidth]{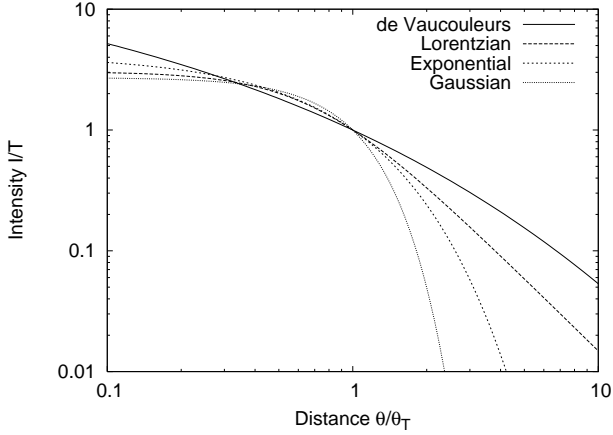}
\end{minipage} \hfill
\caption{The radial profiles of a simulated $V \approx 25.5$ source using the de
Vaucouleurs, Lorentzian, exponential and Gaussian fits. The total intensity within
$\theta_{T} \approx 0.1''$ is constant. The de Vaucouleurs tail is the largest, but it is
small at pair separation scales of $\theta_{min} \approx 8 \theta_{T}$.}\label{fig:PP}
\end{figure}

In principle, the variation of profile brightness at large radius could influence pair
finding. However, for $25 < V < 26$ sources we only search for pairs on scales $\theta >
0.8''$. The average de Vaucouleurs intensity for simulated sources at this distance is
$2.3 \times 10^{-4} \rm{s}^{-1}$ while the GOODS noise threshold is $4.2 \times 10^{-3}
\rm{s}^{-1}$. This value is smaller for all dimmer sources at their respective minimum
pair distance (see section 6). The influence on pair-finding is largest for the de
Vaucouleurs profile that we have chosen to use and even here it does not greatly affect
our final results. Intensities of $5\%$ of the threshold will have minimal influence on
the detection. We plot the intensity of idealized de Vaucouleurs, Lorentzian, Exponential
and Gaussian profile source with $V \approx 25.5$ and typical detection area in Fig.
\ref{fig:PP}. All other profiles drop off faster than the de Vaucouleurs profile on the
pertinent distance scales. The tails at such scales must be minimally important.

\subsection{Simulated Source Distribution}

Our simulated catalogs are designed to match the magnitude, detection area and
ellipticity distributions of GOODS and the UDF. We produced many simulated images using
different input distributions until the output catalogs matched the data.

We start by using a broken exponential magnitude distribution in the \textit{V} band:
\begin{equation}
P(V) \propto \Theta(V-V_{max})\ e^{\eta V}\\
\end{equation}
\begin{eqnarray}
\eta & = & 0.92, \rm{for}\ V < 27.5\nonumber\\
     &  & 0.72, \rm{for}\ V > 27.5
\end{eqnarray}
where $V_{max}$ = $29.5$ for GOODS and $31$ for UDF.

When determining the width of a source, we scale each source so that it can be seen above
the background intensity threshold, T. In the absence of a PSF or ellipticity our sources
would have the de Vaucouleurs intensity profiles:

\begin{equation}
I(r) = \frac{I_0}{8!\ \pi r_0^2 }\ e^{-\left(\frac{r}{r_0}\right)^{1/4}}
\end{equation}
where $I_0$ is just the total integrated intensity of the source.

The term $r_0$ determines the width of our sources. We must make a series of
variable transformations to produce a distribution of input $r_0$'s that leads to an
accurate distribution of detected output widths. We start by noting that for an appropriately
bright and small source, there is some $r_T$ such that $I(r_T) = T$. The size of the
source that we are interested in is the area that exceeds the threshold, $\pi\ r_T^2$. We
define a normalized area above the threshold, $A_T$, and a normalized area $A_0 \propto
r_0^2$. Relating the two tells of how to relate the input source width to the detected
source area:
\begin{equation}
T = \frac{I_0}{8!\ \pi r_0^2 }\ e^{-\left(\frac{r_T}{r_0}\right)^{1/4}}\\
\end{equation}
\begin{equation}
A_0 \equiv \frac{8!\ T\pi r_0^2 }{ I_0}= e^{-\left(\frac{r_T}{r_0}\right)^{1/4}}
\end{equation}
\begin{equation}
A_T \equiv \frac{8!\ T \pi r_T^2}{I_0} = A_0 \log^8(A_0)
\end{equation}

$A_T = 0$ when $A_0 = 1$, because sources more diffuse than this have central
luminosities below threshold. We only make sources with $0\ <\ A_0\ <\ 1$. This function
has sharp, undesirable behavior near its minimum at $A_0 = e^{-8}$. Using the variable
transformation $\alpha = \log(A_0)$ we can make a smoother, manageable function which we
Taylor expand around the maximum at $\alpha = -8$.
\begin{equation}
A_T = \alpha^8\ e^\alpha \approx e^{-8}8^8\left(1-\frac{1}{16}(\alpha+8)^2\right),
\end{equation}
which we can invert to determine a range on the parameter $\alpha$:

\begin{equation}
\alpha_T = -8 \pm \sqrt{16 - \frac{2\ e^8}{8^7} A_T}
\end{equation}

At least 10 pixels must be above threshold or $\pi r_T^2 \geq 10$. This yields a range on
$\alpha$ for which $A_T > 10 \frac{8! T}{I_0}$:

\begin{equation}
-8 - 4 \sqrt{1 - 72 \frac{T}{I_0}} < \alpha < -8 + 4 \sqrt{1 - 72 \frac{T}{I_0}}
\end{equation}

We ignore sources with $I_0/T < 72$ as they are below our detection threshold. After much
experimentation, we find that we can reproduce the actual area distribution of sources
best if we select alpha from a uniform distribution in the bounds:

\begin{equation}
-9 - 4 \sqrt{1 - 115 \frac{T}{I_0}} < \alpha < -9 + 4 \sqrt{1 - 115 \frac{T}{I_0}}
\end{equation}

Finally, we make our sources ellipses with random orientations. The intrinsic ellipticity
is chosen as uniform between 0 and 0.9 which reproduces the ellipticity distribution
after processing. The position of each source is uniform and random except for the
`partner' sources described in subsection \ref{subsec:clustering}.

\subsection{PSF Simulation}

After producing idealized sources we must use an accurate PSF so that small sources are
properly blurred and small angle correlations resemble those of the actual image. We
cannot use a separate PSF for the roughly 120,000 sources in each of several hundred
simulations, so we use a single PSF over our entire field and convolve the simulated
image using the FFTW algorithm \citep{FFTW2005}.

The PSF is constructed using the `Tiny Tiny' program designed to simulate HST PSFs
\citep{KRIS1995}. To simulate an average galaxy in the GOODS fields, we assume a power
law source spectrum with spectral index $n = -0.1$. We average together 100 PSFs at
random positions on each of the two ACS chips and include $0.007''$ of jitter. We apply
the electron diffusion PSF in the Tiny Tim documentation taking care to modify the pixel
size to $0.03''$. The final PSF has a fitted width that is roughly 0.98 times that of the
average fitted width of a random sample of 10 point-like sources.

\subsection{Simulated Noise}

In all simulated images, we use background noise that approximates the small scale
structure of our image and matches the gross statistical properties of the noise in our
data. The noise is a field of Gaussian random numbers with variance proportional to the
inverse of a weight image at each pixel. To simulate drizzling and cosmic correlation, we
convolve this noise with a modified top hat function with bin values proportional to the
fraction of their area that a $0.98$ pixel radius circle would fill. We scale the
standard deviation of the Gaussian field so that we match the roughly $0.0025
\rm{sec}^{-1}$ rms calculated by Oxtractor in GOODS. Our RMS in UDF is roughly $0.0007
\rm{sec}^{-1}$.

With these constraints, the number of counts in the simulated negative image (image
multiplied by $-1$ so that sources are ignored by Oxtractor) is equal to counts in the
negative image when we lower the detection threshold to $1.4 \sigma$ (to increase counts
to roughly 200). Zodiacal light, sunlight scattered off of dust, is the largest
background for HST observations \citep{BERN2002}, and our background is roughly
consistent with a constant zodiacal glow.


\subsection{Comparison of Simulated Image Catalogs with Data Catalogs}

\begin{figure*}
\begin{minipage}[t]{0.49\linewidth}
\centering\includegraphics[width=\linewidth]{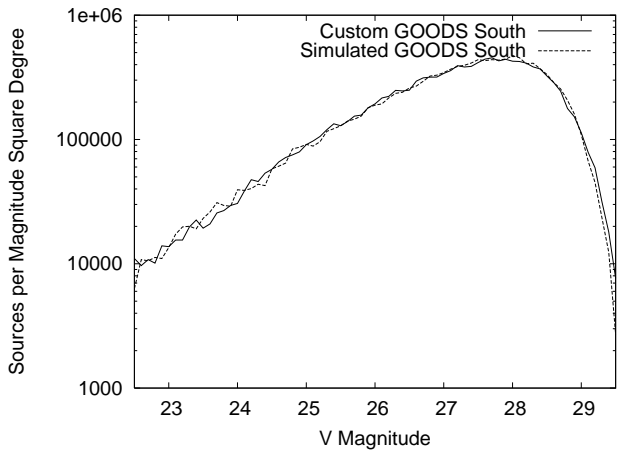}
\end{minipage} \hfill
\begin{minipage}[t]{0.49\linewidth}
\centering\includegraphics[width=\linewidth]{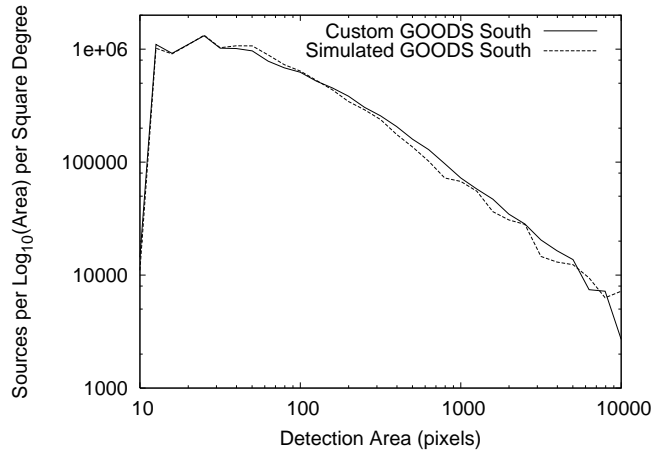}
\end{minipage}
\caption{The magnitude (left) and area (right) distributions of actual and simulated
GOODS data.}\label{fig:comp}
\end{figure*}

The above procedures produced catalogs with similar distributions in \textit{V} magnitude,
detection area and ellipticity as those in the actual GOODS and UDF catalogs as shown in
Fig. \ref{fig:comp}. The number of sources within any magnitude band was within a few
percent of the observed value, and that the area and ellipticity distributions are
similarly close to those observed in the GOODS field.

\subsection{False Detections in Simulated Images}\label{subsec:false}
The detection algorithm in section 4 was guided by our study of false detections in
simulated images. To find false sources, we compared the positions of detected sources to
those of actual sources in our input image. We made a catalog of detected sources that
were more than $0.3''$ (ten pixels) away from actual sources (as defined by an input
catalog of sources). We never include pairs closer than this in our correlation function
calculation. These potential detection areas cover a total of around 10 percent of the
detection area. Increasing the distance at which we are willing to associate a detection
with an input source beyond $0.3''$ decreased the number of false detections at a rate
consistent with the decrease in area that was ``far from a source''. This indicates that
false detections are not strongly clustered around sources on scale greater than
$0.3''$. We find roughly 270 false sources in GOODS South, $0.5\%$ percent of our total
sources. This fraction is roughly constant across magnitude.

\subsection{Simulated Clustering}\label{subsec:clustering}

In order to evaluate our ability to measure an intrinsic correlation function, we must
see how well we measure the correlation function in simulated clustered datasets. To make
these datasets we start with an unclustered data set and assign each source $n_{p}$
partners where the distribution of $n_p$ is:

 \begin{equation}
P(n_p) = \frac{e^{-n_p/n_0}}{n_0};\
n_0 = \rho_0\int_{0.2''}^{10''}\left(\frac{\theta}{\theta_0''}\right)^{-2.5}2\pi\theta
d\theta
\end{equation}
where $\theta_0 = 0.432''\ (0.27'')$ for GOODS (UDF).

These clustered sources increases the total number of sources by $50\%\ (33\%)$. Each
extra source is assigned a separation angle, $\theta$,  from its parent with a
distribution:
 \begin{equation}
P(\theta) \propto \theta^{-1.5}
\end{equation}
Between a minimum $\theta$ of $0.2''\ (0.1'')$ in GOODS (UDF) and a maximum $\theta$ of
$10''$.We pick a uniformly distributed random position angle $\phi$.

This method produces power law distributed clumps but because of clump-clump correlations
does not produce perfect power law behavior. Nor does it produce an exact match to the
observed correlation function. We use these simulations only to study how well our
measurement algorithm recovers an intrinsic correlation.

\section{Catalog Production}

Attempts to measure $w(\theta)$ and $\xi(r)$ in bright galaxy surveys are rarely confused
about what is being counted. Bright $r < 22$ galaxies are physically distinct 'island
universes', and although they are observed to collide and merge, the autocorrelation
statistics are not seriously hindered by decisions about whether or not to count a
comparatively rare interacting pair as one galaxy or two. However, when considering the
FSCF in our sample, we quickly realize that $\theta_0$ is only a few times larger than
the physical size of the sources and the resolution of the observations. This implies
that we must be scrupulous in defining sources and consistently use the same definition
when comparing with simulations of galaxy formation.

\subsection{Source Extraction}

When making our catalogs for FSCF study, we designed a source extraction routine geared
to  look for faint, compact sources and deblend aggressively. In exchange for this
increase in performance, we allowed for around $0.5\%$ false sources that the more
conservative GOODS catalog lacks.

We started with the catalog procedures used by the GOODS team and modified them to look
for faint sources and pairs. In making their catalogs, the GOODS team used a modified
version of the SExtractor \citep{BERT1996} program called Object Extractor (Oxtractor)
that is designed to better extract faint sources near bright neighbors by modifying the
noise floor in these areas. Oxtractor also avoids including spurious noise in the area of
a source as part of that source. We borrowed their code to produce our own catalog.

We modified the GOODS team's procedures at several stages. Our most distinct change in
method from the GOODS team was to use the \textit{V} band (F606W)instead of the z band (F850LP).
Using any reasonable SExtractor parameters designed to find faint sources, we find more
GOODS sources in the \textit{V} band. For our particular set of parameters, we found 56088 source
in \textit{V} band and only 29601 in z. This suggests that many faint source are many blue star
forming regions. We also changed the actual Oxtractor detection parameters to better find
faint, small sources. Our changes are summarized in Table \ref{tab:sexchange}. One should
note that the GOODS team used RMS images (not publicly available) that are normalized
differently from our weight images and that, accounting for this difference, our
DETECT\_THRESH is roughly equivalent to theirs.

\begin{table}
\begin{tabular}{ccc}
                & Standard GOODS Catalog & Custom Catalog\\
\hline
WEIGHT\_TYPE        & MAP\_RMS & MAP\_WEIGHT\\
Filtering FWHM (Pixels) & 5.0 & 1.5        \\
DETECT\_THRESH        & 0.6 & 1.7        \\
DETECT\_MINAREA        & 16 & 10        \\
DEBLEND\_NTHRESH & 32 & 16        \\
DEBLEND\_MINCONT & 0.03 & 0.03        \\

\hline
\end{tabular}
\caption{The standard and custom GOODS SExtractor parameters. }\label{tab:sexchange}
\end{table}

We arrived at these numbers by examining the correlation function of the $27<V<28$
sources in uncorrelated simulation images, the number of detections in real images and
the number of false counts in simulated images. To determine a filtering FWHM, we ran
Oxtractor with different Gaussian filters of width between 1 and 5 pixels
($0.03''-0.15''$) to produce catalogs from simulated uncorrelated data. The correlation
function of these catalogs would ideally be zero for all values of $\theta$, but we found
that it became significantly negative at distances of roughly 3 times the FWHM. Running
Oxtractor without filtering or while using only a small filter causes the correlation
function to become very large on scale less than $0.2''$. A FWHM of $1.5$ pixels was the
smallest we could use without introducing this small angle peak.

We again used this correlation function to study the DEBLEND\_NTHRESH-DEBLEND\_MINCONT
parameter space. We qualitatively found the parameters which minimize spurious
deblending, manifested by a large correlation of what should be uncorrelated data at
separations of roughly $0.1-0.3''$, without hindering our effective resolution,
manifested as a suppression of $w(\theta)$ on scale of $0.2''-0.5''$. Our final choice
was identical to \citet{BENI2003}.

We examined the 2 dimensional space of DETECT\_MINAREA and DETECT\_THRESH using total
counts from the real image and false counts in a simulated image. The PSF is roughly 3
pixels wide, so we centred our search around DETECT\_MINAREA $\approx 9$. We tried every
integer value between 6 and 20 pixels. Roughly speaking, we wanted to focus on aggressive
$5 \sigma$ detections. For DETECT\_MINAREA $= 9$, a $5 \sigma$ detection corresponds to
DETECT\_THRESH $\approx 1.7$. We tested every $0.1$ interval of DETECT\_THRESH between
0.6 and 3. Our goal was to obtain the largest number of detections with $99.5\%$ purity.
Our final setting of 10 pixels at 1.7 yields an average of 60,000 detections in a
simulated GOODS South data set with 270 false detections.

\begin{figure}
\centering\includegraphics[width=\linewidth]{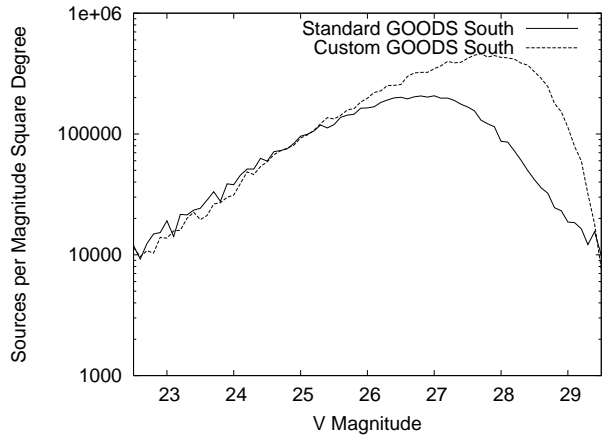}
\caption{The magnitude distribution in the standard and custom GOODS South catalogs. The
custom GOODS catalog contains nearly twice as many sources. }\label{fig:magdist}
\end{figure}

Our source extraction procedure detects only 51 negative sources in GOODS South and an
average of about 270 false detections in simulated images (as described in section 3).
This catalog raises the total number of counts from 29601 to 56088 in GOODS South. In
addition, Fig. \ref{fig:magdist} shows that we improved completeness of extended
sources from roughly \textit{V} = 26 to V= 27.5. To justify our previous claim that the \textit{V} band is
preferable for detecting many faint sources, we note that we detect only 29488 sources
and 67 negative sources on the z band with the above parameters.

\subsection{Masking}\label{subsec:masking}

Our source extraction methods fail in two areas of our images, so we masked these areas
out separately. High noise near the edges of our images produce a large number of
correlated pairs. Improper background subtraction and real structure near bright sources
produced many suspect pairs.

Near the edges of our images, effective exposure time drops off, and the background noise
becomes large. These areas looked qualitatively different from the rest of the image and
produced a disproportionate number of close pairs, so we removed them from our images. To
do this, we convolved the weight images with a 30 pixel ($0.9''$) top hat and excised all
areas in the original weight and science images where the weight in this convolved image
was less than 40,000. These masked area amounted were concentrated almost entirely near
the edges of the images and amounted to roughly $2\%$ of the original images.

Bright ($V < 21$) sources provide two separate problems. First, errors in background
subtraction from these sources affect significant area and are not accounted for in our
source extraction procedures. Second, our aggressive deblending means that a small number
of bright, nearby sources with complex structure can be split into many faint pairs and
significantly influence the overall correlation function. We are only interested in
studying faint sources that are not obviously associated with a bright source. To mask
out these bright sources, we reject all sources contained in ellipses which are four
times the area of the bright source. These masked areas amounted roughly $2\%$ of the
original images.

\subsection{Characterizing the Catalogs}

\begin{figure*}
\begin{minipage}[t]{0.32\linewidth}
\centering\includegraphics[width=\linewidth]{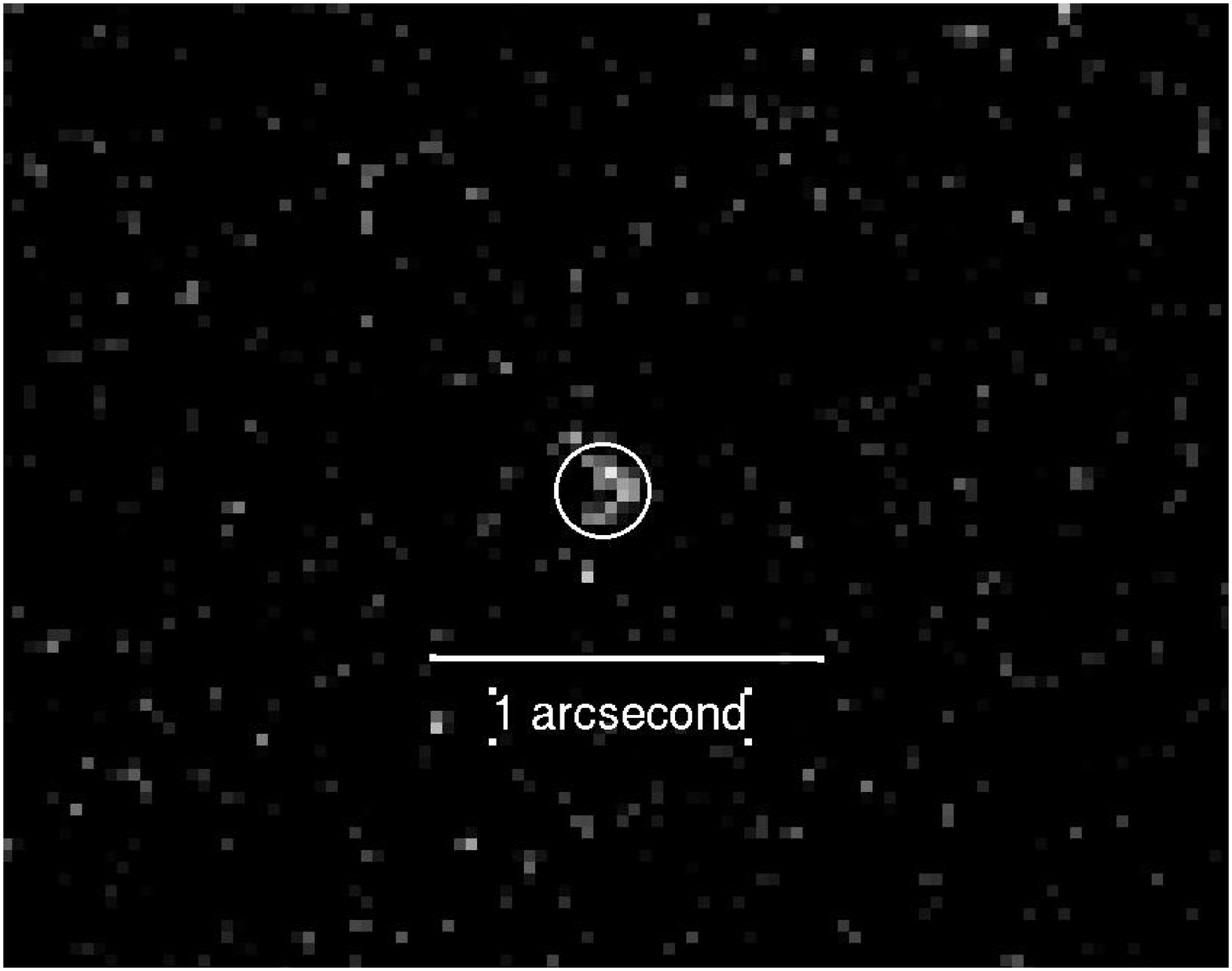}
\end{minipage}
\begin{minipage}[t]{0.32\linewidth}
\centering\includegraphics[width=\linewidth]{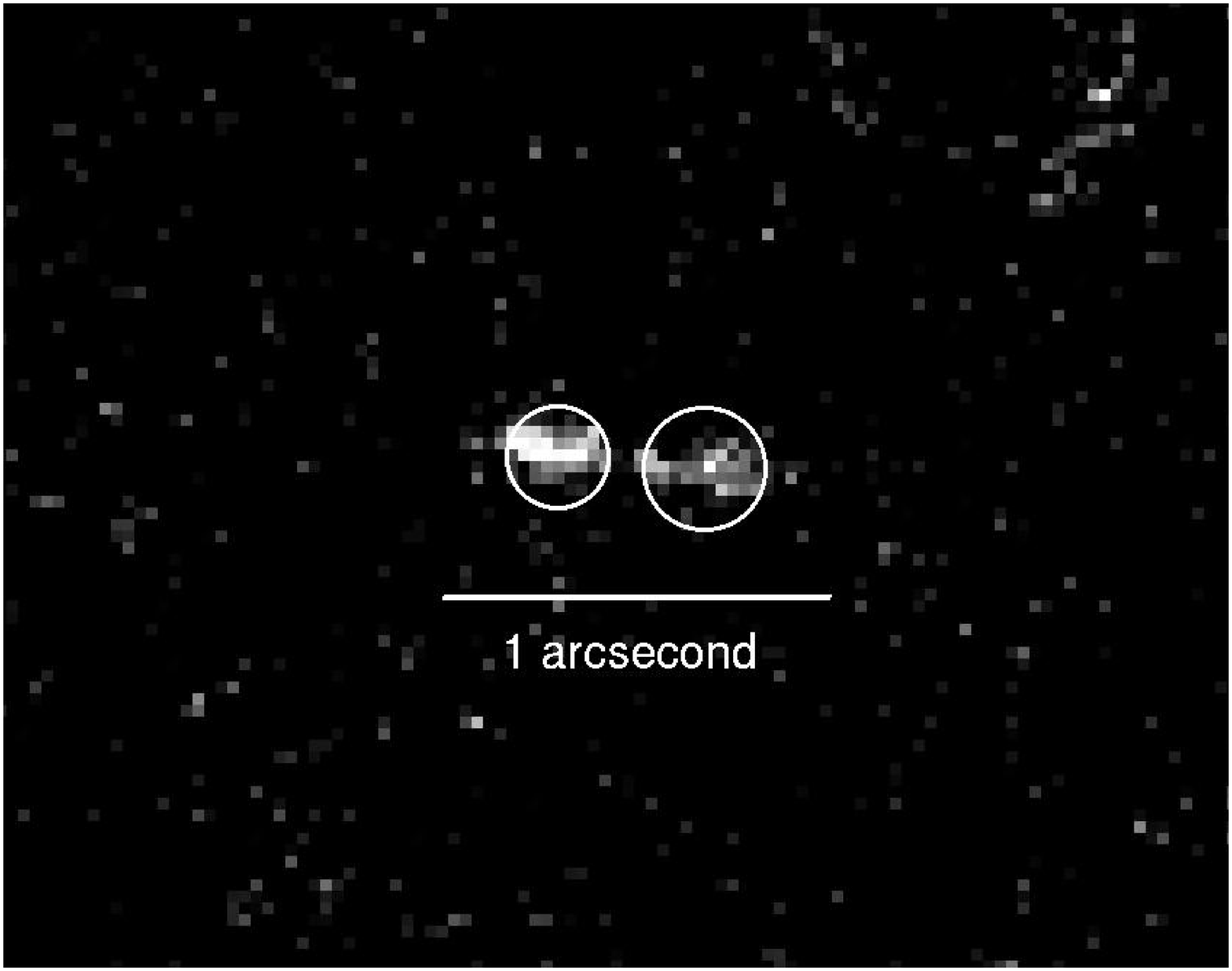}
\end{minipage}
\begin{minipage}[t]{0.32\linewidth}
\centering\includegraphics[width=\linewidth]{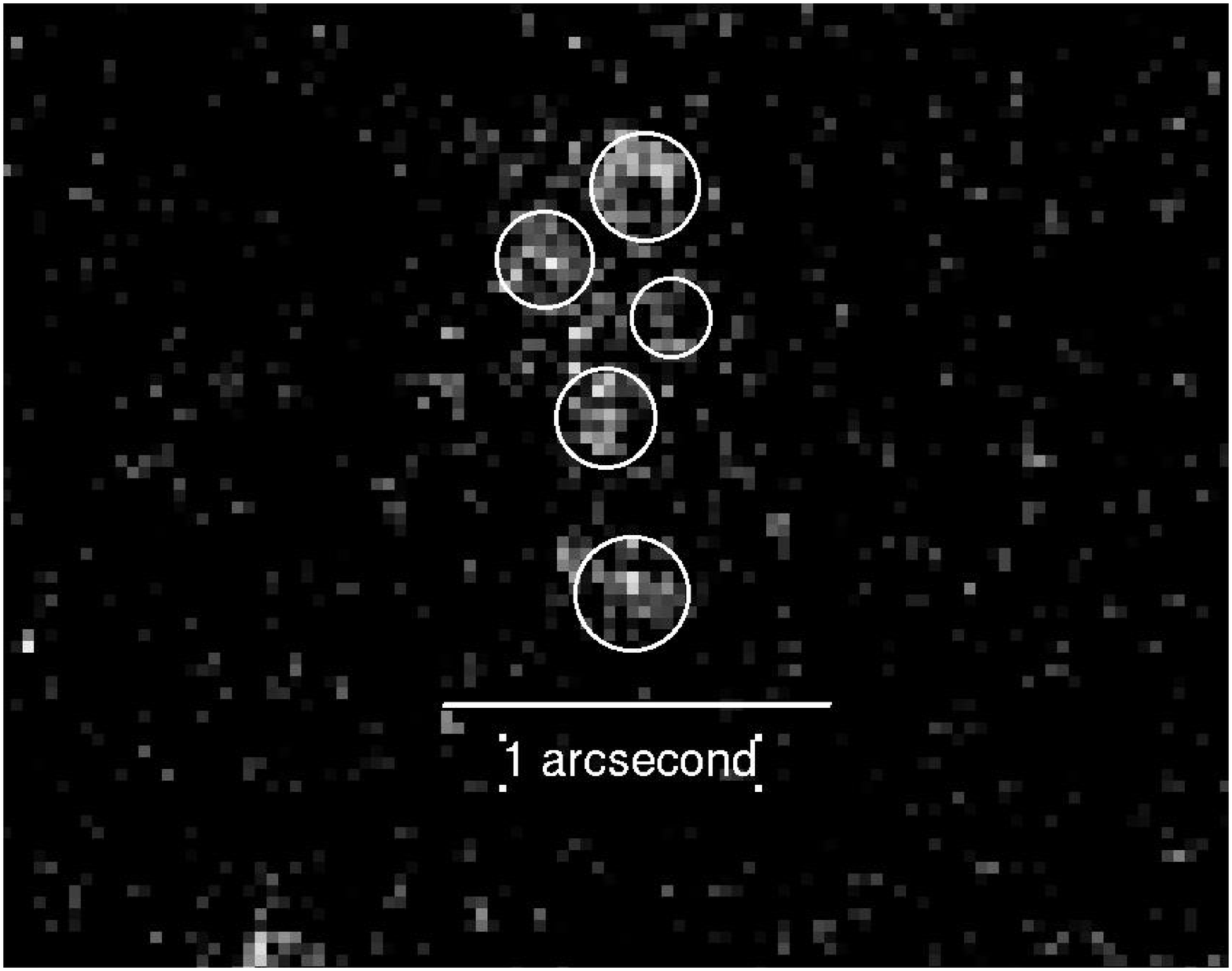}
\end{minipage}
\caption{A typical singleton, pair and cluster of faint source. These sources are from
the GOODS South field and range in magnitude between $V = 26.3$ (the second lowest source
in the cluster) and $V = 27.5$ (the singleton).}\label{ds9}
\end{figure*}

Our catalogs contain many faint sources that are small in extent and separation. The
example sources in Fig. \ref{ds9} range in magnitude between $V = 26.31$ and $V = 27.49$
and in diameter between $0.1''$ and $0.3''$. The separation lengths between detected
sources range between $0.3''$ and $1.2''$.

These are $3.5 - 2.5$ magnitudes dimmer than those used in large scale studies like SDSS.
Converting from SDSS r band to HST \textit{V} band is an imprecise technique. We observe that
sources have $B-V$ of roughly 1.1 and make the rough conversion between detection bands
via \citep{JEST2005}:
\begin{equation}
r = V - 0.42 (B-V) + 0.11 \approx V-0.36
\end{equation}

Given the approximate nature of the spectra, this should be taken as only a rough
conversion, but the SDSS cutoff magnitude of $r = 22.5$ is roughly equivalent to a
GOODS/UDF $V = 22.9$ cutoff, two magnitudes brighter than the dimmest sources we use.

We also work on much smaller scales. Typical source separation in SDSS were $100''$,
resolution limits source size to $1.4''$ and statistics limited their source separation
to $10''$. Our typical source separation is roughly $5''$, our resolution is $0.12''$ and
we see pairs with $0.3''$ separation.

We place these numbers in an astrophysical context in table \ref{table:local}. For
reference, M31 has an absolute magnitude of roughly $V = -20$ and the bulk of its
luminosity is from a disk roughly $20$ kpc across.

\begin{table}
\begin{tabular}{ccccc}
        z        & Absolute \textit{V} & Arcsecond Linear Size (kpc) \\
\hline
0.1 & -11.3 & 1.8\\
0.2 & -12.9 & 3.3\\
0.5 & -15.3 & 6.1\\
1   & -17.1 & 8.0\\
2   & -19.0 & 8.5\\
5   & -21.4 & 6.4\\

\hline
\end{tabular}
\caption{The absolute magnitude of a source with apparent magnitude 27 (in an
appropriately blueshifted \textit{V} band) and physical distance corresponding to $1''$ at various
redshifts.\ref{ds9}.}\label{table:local}
\end{table}

Finally in Table \ref{table:stats}, we note the number sources, $10''$ pairs and $1''$ pairs. The crucial number is the number or $1''$ pairs. The Poisson noise of this number gives us a rough idea of how well we can measure our subarcsecond correlation function and shows that we cannot avoid at least a few percent error.

\begin{table}
\begin{tabular}{ccccc}
        Sample        & Subsample & Sources & $10''$ pairs & $1''$ pairs \\
\hline

GOODS South & $25 < V < 26$ & $6266$  & $10653$ & $591$ \\
            & $26 < V < 27$ & $12282$ & $41049$ & $1514$ \\
            & $27 < V < 28$ & $18356$ & $92252$ & $1885$ \\

GOODS North & $25 < V < 26$ & $6535$  & $11405$ & $626$ \\
            & $26 < V < 27$ & $12080$ & $36603$ & $1442$ \\
            & $27 < V < 28$ & $18016$ & $81140$ & $2015$ \\
UDF         & $27 < V < 28$ & $1339$  & $6154$  & $141$ \\
            & $28 < V < 29$ & $2111$  & $15394$ & $227$ \\
\hline
\end{tabular}
\caption{The number of sources, $10''$ pairs and $1''$ pairs for each subsample.\ref{ds9}.}\label{table:stats}
\end{table}

\section{Correcting Measurement Error in the Correlation Function}

For large angle correlation functions, nonuniform survey depth is usually the major threat of
measurement error. After masking out the edges of the survey
(see subsection \ref{subsec:masking}) we have relatively uniform surveys and particularly
very little survey depth structure on the arcsecond scale. In addition, conspicuous
causes of `false sources' such as diffraction spikes and cosmic rays near the edge of
the image where drizzling is not effective are again cleanly removed by this masking.

Proper deblending of distinct objects and improper deblending of single source is our
main source of measurement error. We examine this problem from two perspectives. First,
we estimate a correlation function caused by imperfect measurement techniques using
intrinsically unclustered simulated data. Secondly, we estimate our ability to measure
actual clustering using simulated clustered data.

\subsection{Measurement-Induced Correlations in Uncorrelated Data}

Our approach is an extension of the statistical method for evaluating the correlation
function introduced by \citet{HAMI1993}. Specifically, to minimize the effects of
nonuniform weighting and uncertainty in the zero point of $w(\theta)$ we use a
combination of two of Hamilton's estimates of the correlation function:
\begin{eqnarray}
1+w_{est}(\theta) & = & \frac{<DD> <RR>}{<RD>^2 } \\
1+w_{est}(\theta) & = & \frac{<DD>}{<RR> }
\end{eqnarray}

Here, $<DD>$ is the simple estimate of the data-data correlation function taken as a
series of delta functions representing each pair separation. $<RR>$ is the random-random
correlation function of many (in our case 40) random fields emulating the appropriate
dataset, and $<RD>$ is the correlation of the appropriate data field with its
corresponding random fields. We convolve $<RR>$ and $<RD>$ with a Gaussian filter with
angular width $0.1''$ to produce continuous functions. We normalize each function by a
factor of roughly $\sigma_0^2$ such that it converges to unity at large angles.

The first method excels at reducing error due to survey nonuniformity. But at small
angles, resolution and source size reduce $<DD>$ and $<RR>$, but not $<RD>$, and the
estimate  of $w(\theta)$ is artificially damped. The second method does not counter
survey nonuniformity as well as the first, but the damping in $<DD>$ and $<RR>$ cancel to
first order to give a more accurate measurement at small angles. Therefore, we use the
first method for $\theta \geq 2''$ and the second method for $\theta < 2''$.

\begin{eqnarray}
1+w_{est}(\theta)& = & \frac{<DD> }{1+w_M(\theta)}\\
1+w_M(\theta) & = & \frac{<RD>^2}{<RR>}, \rm{\ for\ }\theta > 2''\nonumber\\
              &  & <RR>, \rm{\ for\ }\theta < 2''
\end{eqnarray}
where we have defined $w_M$, the measurement-induced correlation function, for our own
convenience.

\begin{figure*}
\begin{minipage}[t]{0.49\linewidth}
\centering\includegraphics[width=\linewidth]{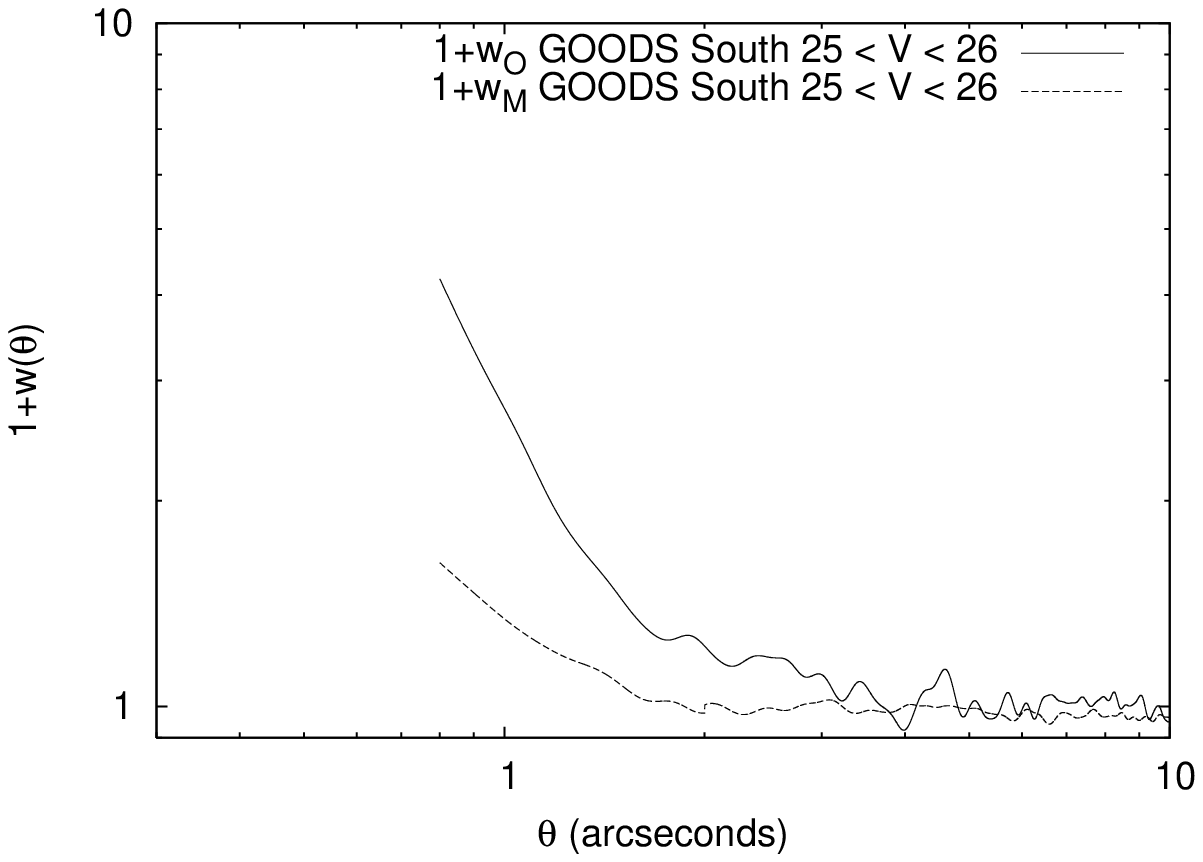}
\end{minipage} \hfill
\begin{minipage}[t]{0.49\linewidth}
\centering\includegraphics[width=\linewidth]{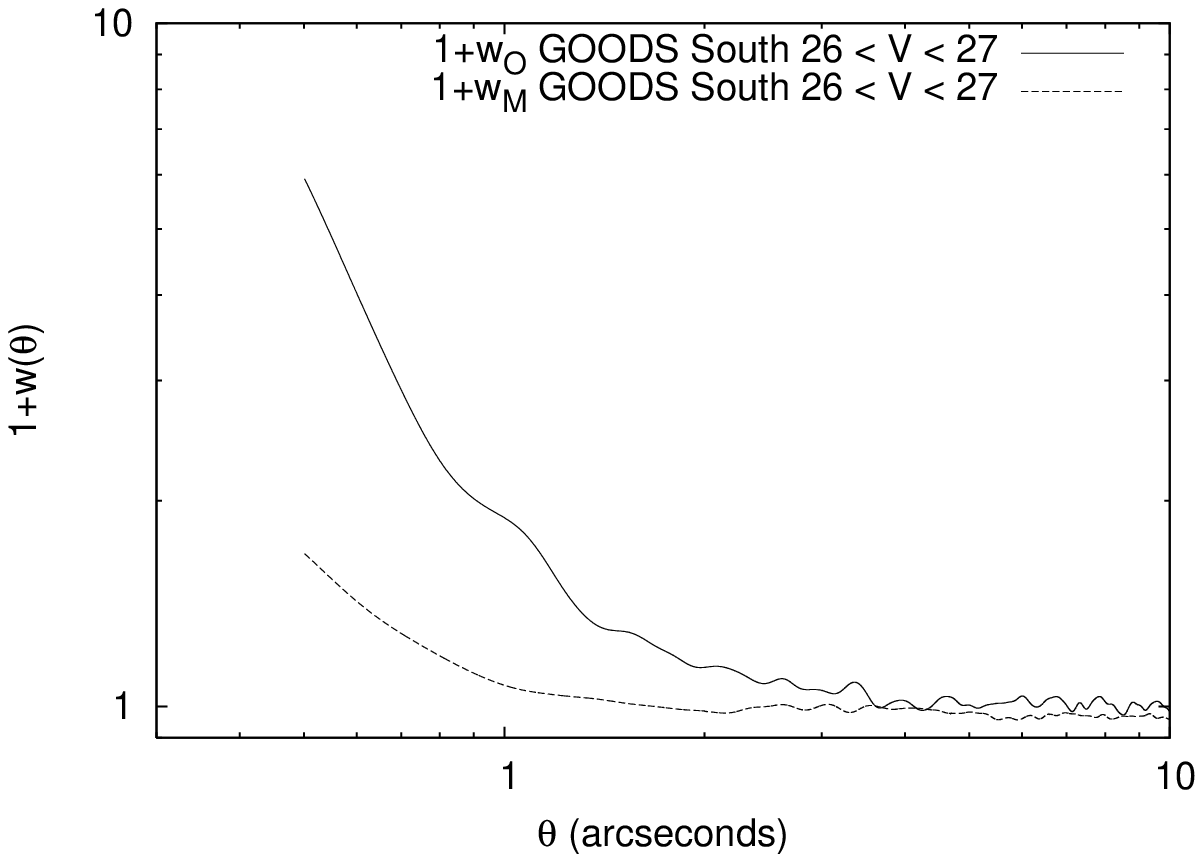}
\end{minipage}
\begin{minipage}[t]{0.49\linewidth}
\centering\includegraphics[width=\linewidth]{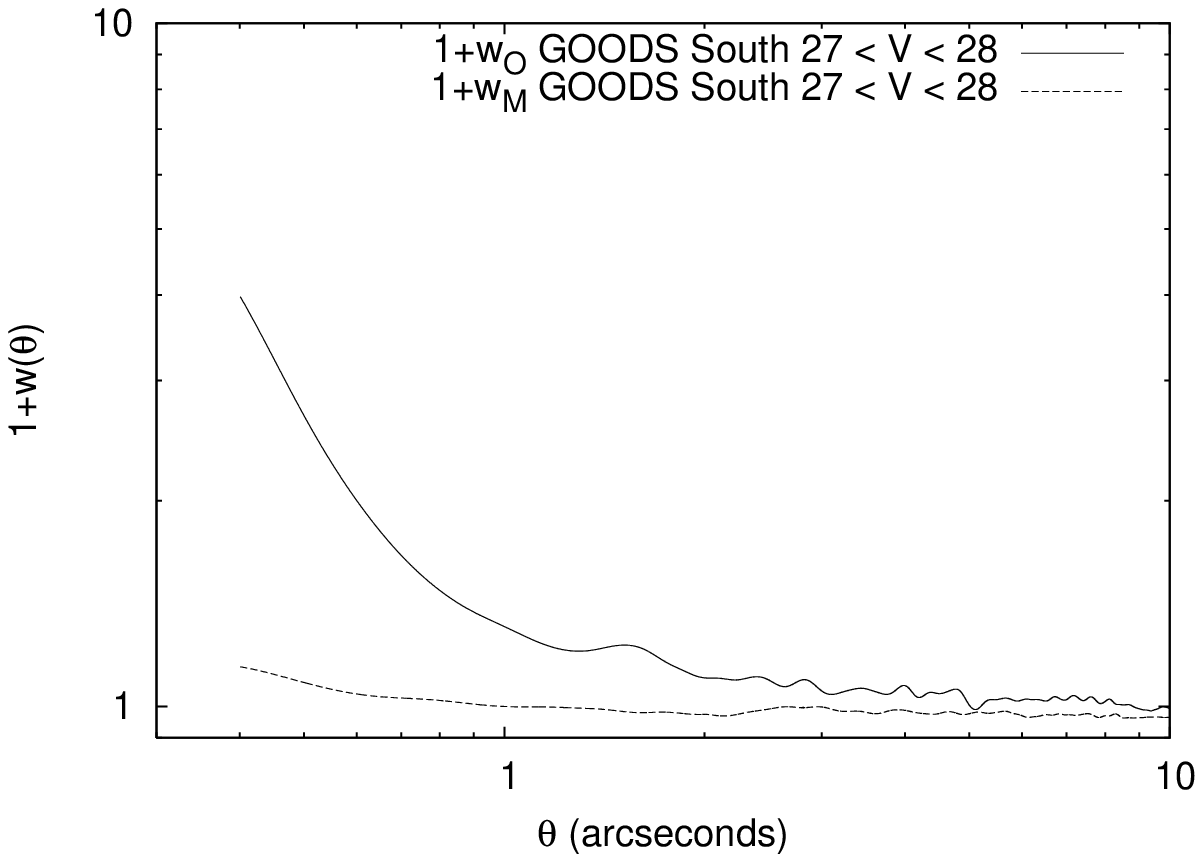}
\end{minipage}
\begin{minipage}[t]{0.49\linewidth}
\centering\includegraphics[width=\linewidth]{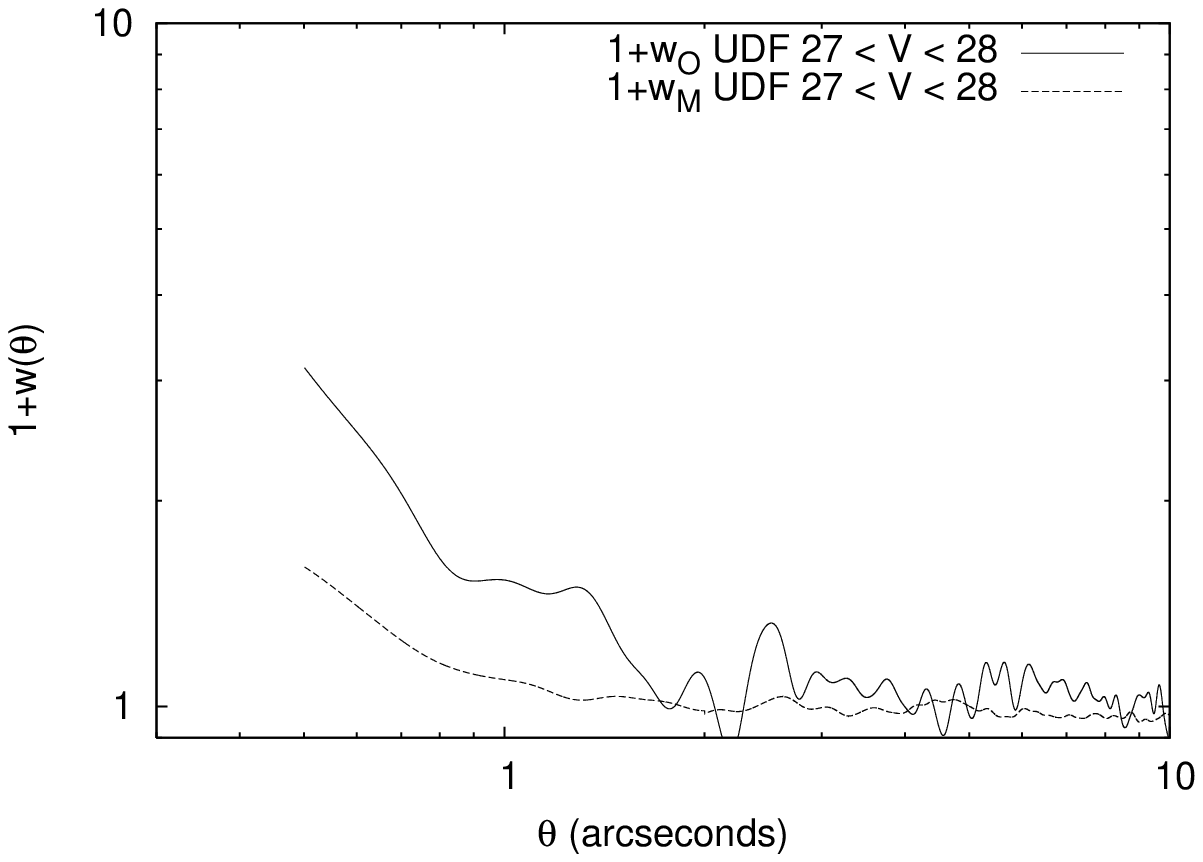}
\end{minipage}
\begin{minipage}[t]{0.49\linewidth}
\centering\includegraphics[width=\linewidth]{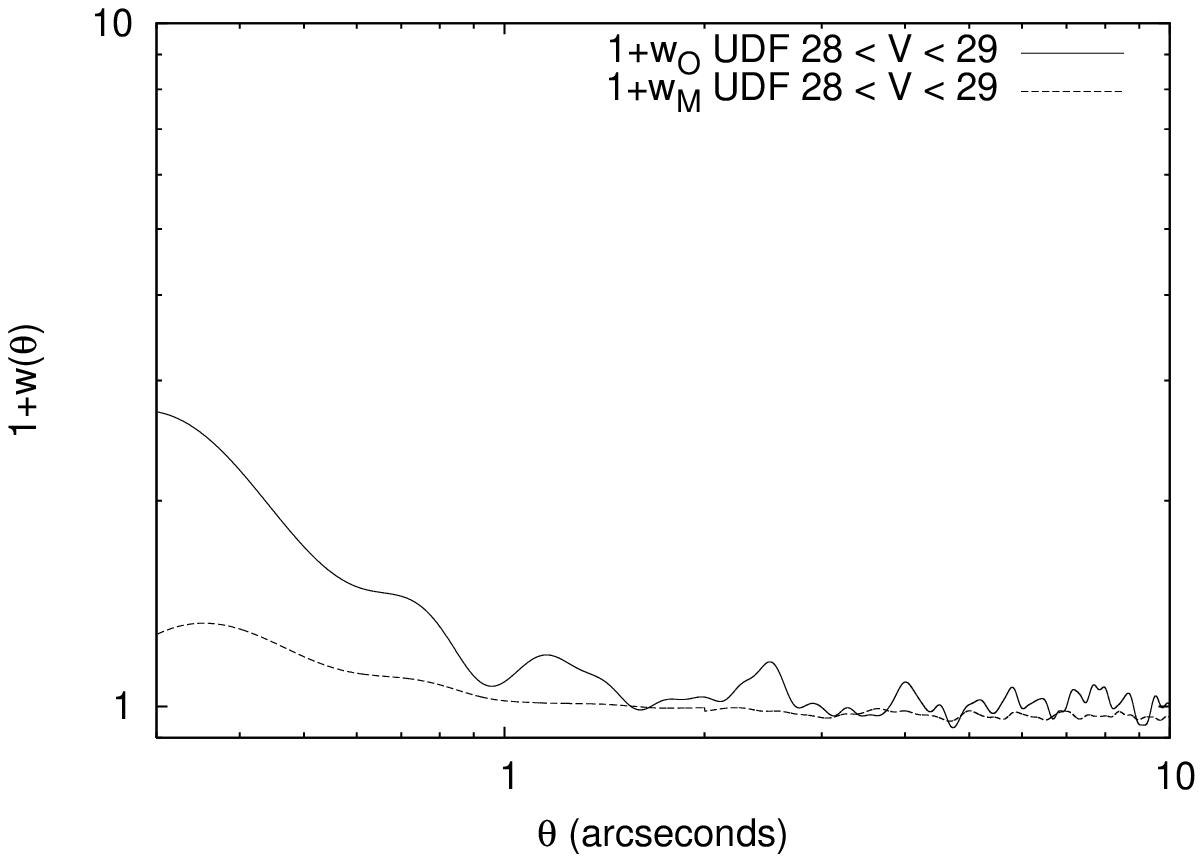}
\end{minipage}
\caption{We compare $w_O$, the naive observed correlation function in our data and $w_M$, the correlation function induced on a random field of sources via
spurious deblending, in the GOODS South field and the UDF.}
\label{fig:wsys}
\end{figure*}

To first order, $w_M$ is the correlation function one would observe from a naive $<RR>$
measurement of uncorrelated sources due to survey geometry and improper deblending.  This function can be positive if a single source is improperly deblended or if dim sources are enhanced by the tail of a bright source. It can be negative if two sources are separated by less than their angular extent or the resolution of the instrument. In Fig. \ref{fig:wsys}, we compare the observed data-data correlation function $ 1+w_O = <DD>$ and $1+w_M$ (each convolved with a $0.1''$ filter). $w_M$, is always smaller than $w_O$, but on small scales, it is generally within an order of magnitude of $w_M$ and must be handled intelligently to prevent significant systematic error.



\begin{table}
\begin{tabular}{c|cc}
                &GOODS & UDF\\
\hline
$25<V<26$        & $0.8$  & ---  \\
$26<V<27$        & $0.55$ & ---\\
$27<V<28$        & $0.4$  & $0.5$  \\
$28<V<29$        & ---    & $0.3$  \\
\hline
\end{tabular}
\caption{$\theta_{min}$ in GOODS and the UDF. }
\end{table}

We use our estimation of the measurement correlation function in Fig. \ref{fig:wsys}
to choose $\theta_{min}$
below which we do not study the correlation function. We define $\theta_{min}$ as roughly
$0.1''$ greater than the angle where $w_{M}$ approaches unity or where it becomes
negative. This corresponds roughly to the source size, below which our simple source
profile assumptions should become important.

\subsection{Correcting Measurement Error}
For large scale correlation functions of bright sources, the naive observed correlation
function, $w_O = <DD>-1$, and the physical correlation function, $w_P \approx w_{est}$,
are simply related:
\begin{equation}
1+w_O = (1 + w_P)(1 + w_M)= 1 + w_P + w_M + w_P w_M
\end{equation}

But because we are measuring the correlation function on scales similar to the PSF and
source size, we must adapt this method to account for these effects. We find that the 
suppression of the correlation on small scales is dependent on the amplitude of the 
correlation function and the magnitude of the sources being measured. We use a fitting 
model that accounts for these dependencies and is accurate well within our statistical 
error bars.

We must relate the known $w_O$ and $w_M$ to an unknown $w_P$. For small values of $w_P$ 
and $w_M$, $w_O$ should equal $w_M + w_P$ since there is only perturbative clustering 
and measurement error. To make a first order approximation when $w_P$ and $w_M$ are not both 
small we use the following model:
\begin{equation}
1+w_O = 1 + w_P + w_M + \lambda w_P w_M
\end{equation}

For the standard correlation function measurement, $\lambda = 1$ to make this a product.
But we find that varying $\lambda$ is a convenient way to account for nonlinear effects
of measurement error in crowded fields. The $\lambda$ term is only significant on small 
scales where both $w_P$ and $w_M$ are large.

In addition, we find that for incomplete samples, we underestimate the correlation function even at large separations. We introduce parameter $\kappa$ to correct for this effect in faint, incomplete samples ($27 < V < 28$ in GOODS and $28 < V < 29$ in UDF):

\begin{equation}
1+w_O = 1 + \kappa w_P + w_M + \lambda w_P w_M
\end{equation}

\begin{table}
\begin{tabular}{c|cccc}
                &GOODS  && UDF&\\
  & $\kappa$ & $\lambda$  & $\kappa$ & $\lambda$ \\
\hline
$25<V<26$        & $1$& $0.8$ & --- & ---  \\
$26<V<27$        & $1$& $0.3$ & --- & ---\\
$27<V<28$        & $0.64$& $0.0$&$1$ &$-0.4$  \\
$28<V<29$        & --- & ---        & $0.89$&$-1.2$  \\
\hline
\end{tabular}
\caption{$\lambda$ and $\kappa$ in used to reproduce input correlation functions in
simulated data. }\label{table:kl}
\end{table}

\begin{figure*}
\begin{minipage}[t]{0.49\linewidth}
\centering\includegraphics[width=\linewidth]{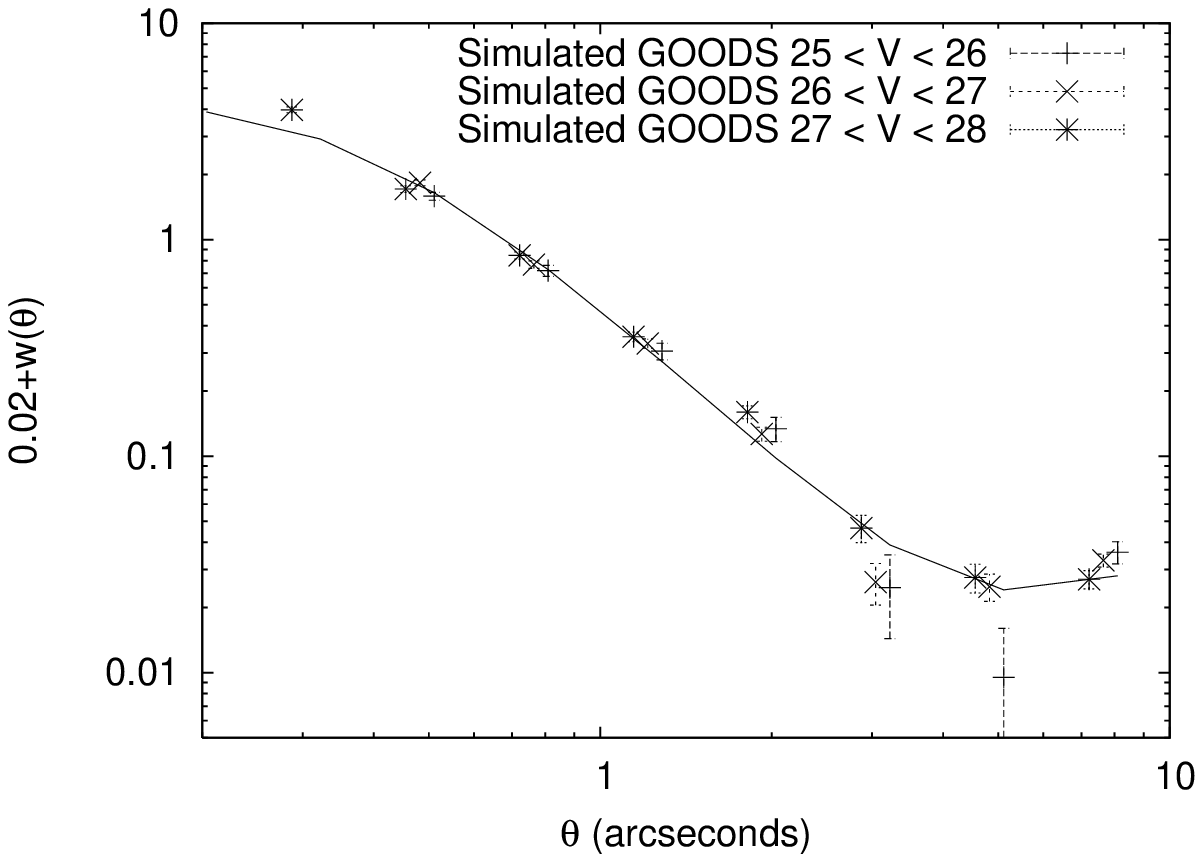}
\end{minipage} \hfill
\begin{minipage}[t]{0.49\linewidth}
\centering\includegraphics[width=\linewidth]{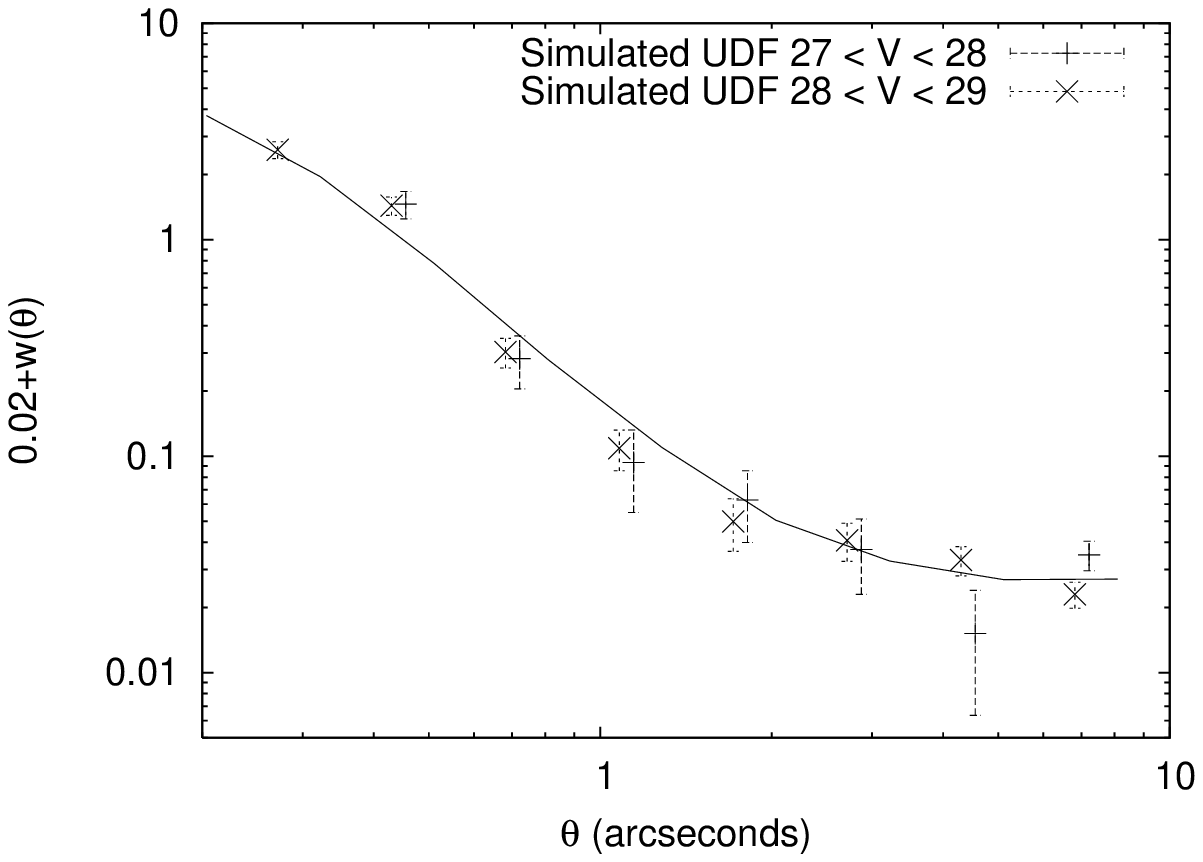}
\end{minipage} \hfill
\caption{ $w_P$ (lines) and $(w_0-w_M)/(\kappa+\lambda\ w_M)$ (datapoints) in simulated
GOODS (left) and UDF(right). We obtain $w_P$ by finding the correlation function in the
input catalogs we use to make our clustered simulated images.}\label{fig:kldiff}
\end{figure*}

Using the $\kappa$ and $\lambda$ parameters, we can reconstruct $w_P$ from $w_O$ and
$w_M$ using:

\begin{equation}\label{convert}
1+w_P = 1 + \frac{w_0-w_M}{\kappa+\lambda\ w_M}
\end{equation}
as shown in Fig. \ref{fig:kldiff} for simulated clustered data. In these plots we show 
$0.02+w$ to facilitate logarithmic plotting. We calculate $w_P$ using
the known positions of sources in our simulated clustered images. The fits in Fig.
\ref{fig:kldiff} guided us to use this parameterization.


Models which related higher powers of $w_P$ and $w_M$  to $w_O$ did not improve our ability to reconstruct the input correlation function significantly. Our simple $\lambda$ fit allowed us to find simulated $w_p$ with precision much greater than the statistical fluctuations in a single GOODS or UDF measurement, so it is sufficient for our purposes. More precise models may be employed for future work. The essential requirement for any such method is that close pair suppression vary with the amplitude of $w_P$.

\section{The Estimated Correlation Function}
We employ a maximum likelihood estimation technique to  measure the correlation function,
$w(\theta)$ without binning in $\theta$. Mathematically, not binning is equivalent to
binning very finely so that each in has either one or zero sources in it. If we assume a
Poisson distribution in each bin, the likelihood of a particular realization would be:

\begin{equation}
  L(\sigma_0, \theta_0, \Gamma) = \prod_{i=1}^{n} e^{-\mu_i} \mu_i^b
\end{equation}
where b is equal to the number of pair correlations within this angle bin (either 0 or 1)
and $\mu_i$ is equal to the expected number of pairs in this bin:
\begin{equation}
  \mu_i = \sigma_0^2 (1 + w_M(\theta_i)+ \left(\theta_i/\theta_0\right)^{-\Gamma}
(\kappa+\lambda w_M(\theta_i))) A\ 2\pi\theta_i\ \delta \theta
\end{equation}
where A is the area of the sample and $\delta \theta$ is the bin width, constant and
small enough so that $\mu << 1$.

Implicitly, we have some maximum and minimum $\theta$ defined as
$\theta_{max}-\theta_{min} = n\ \delta \theta$. Hence:
\begin{eqnarray}
\log(L)& = & \sum_{i=1}^n (-\mu_i + b_i\log(\mu_i)) \\
& = &-\sum_{i=1}^n \mu_i + \sum_{pairs}\delta\theta + \\
& &\sum_{pairs}\log(\sigma_0^2\left(1+(\theta_i/\theta_0)^{-\Gamma}(\kappa +\lambda w(\theta_i))\right) 2\pi \theta_i)\nonumber
\end{eqnarray}

In the continuous limit,
\begin{equation}
  \log(L) =
-\int_{\theta_{min}}^{\theta_{max}}\mu(\theta)d\theta+\sum_{pairs}\log(\mu(\theta))+\rm{constant}
\end{equation}
where we have abbreviated the sum involving only $\delta\theta$ as merely a constant
which will be unimportant in the maximization process.

We introduce a continuous version of $\mu$:
\begin{equation}
  \mu(\theta) = \sigma_0^2 (1 + w_M(\theta)+
\left(\theta/\theta_0\right)^{-\Gamma}(\kappa+\lambda w_M(\theta))) A\ 2\pi\theta\
\end{equation}

We maximize L over $\sigma_0^2$, $\Gamma$ and $w_1 = ( \theta_0 )^\Gamma$. To find the
error bars for each parameter, we perturb its value, maximize over the remaining
parameters and use the second derivative of this marginalize likelihood to estimate $1
\sigma$ errors.

%


\subsection{Fit Values}

We present our final values in the GOODS North field, the GOODS South field and the UDF
in Table \ref{tab:fullresults}. The systematic error bars are explained in subsection
\ref{subsec:sys}. The results are consistent. Particularly, our GOODS results for $27 < V
< 28$ agree with our UDF results for the same sources.

Comparison with previous results is necessarily indirect. \citet{VILL1997} produced the most comparable measurement, but used $r$ band limits rather than $V$ band ranges. As a point of reference, we see that our $25 < V < 26$ measurement and their $20 < r < 26$ are within statistical error bars for $\theta > 1''$. More broadly, we agree that $w(\theta) \approx 0.1$ for $\theta > 1''$ for these faint sources. It is our probing down to $w(\theta)$ for $\theta < 1''$ that gives us a significant measurement.

\begin{table*}
\begin{tabular}{c|ccc}
                & $w(1'')$ & $\theta_0$ &        $\Gamma$ \\
\hline
GN $25<V<26$        & $0.83 \pm 0.05 \pm 0.12$    & $0.93  \pm 0.10  \pm 0.04$  & $2.48
\pm
0.37$ \\
GS $25<V<26$        & $0.92 \pm 0.05 \pm 0.10$    & $0.97  \pm 0.11  \pm 0.04$  & $2.55
\pm
0.15$ \\
GN $26<V<27$        & $0.536 \pm 0.019 \pm 0.053$ & $0.775 \pm 0.043 \pm 0.031$ & $2.45
\pm
0.17$ \\
GS $26<V<27$        & $0.508 \pm 0.020 \pm 0.051$ & $0.759 \pm 0.042 \pm 0.031$ & $2.45
\pm
0.17$ \\
GN $27<V<28$        & $0.36  \pm 0.015 \pm 0.10$  & $0.649 \pm 0.035 \pm 0.074$ & $2.38
\pm
0.14$ \\
GS $27<V<28$        & $0.234 \pm 0.013 \pm 0.067$ & $0.549 \pm 0.033 \pm 0.074$ & $2.42
\pm
0.20$ \\
UDF $27<V<28$        & $0.296 \pm 0.053 \pm 0.029$ & $0.60  \pm 0.10  \pm 0.02 $ & $2.41
\pm
0.63$ \\
UDF $28<V<29$   & $0.087 \pm 0.024 \pm 0.009$ & $0.438 \pm 0.042 \pm 0.016$ & $2.96 \pm
0.50$ \\
\hline
\end{tabular}
\caption{$w(1'')$, $\theta_0$ (in arcseconds) and $\Gamma$ for GOODS North (GN), GOODS
South (GS) and the UDF. }\label{tab:fullresults}
\end{table*}

The GOODS North and South results are in good agreement. So we make a final estimate
combining the two datasets to minimize statistical noise. We then have our best estimate
of the correlation function in each of the four magnitude bins in Table \ref{tab:bestw}.
In the $27 < V < 28$ bin we use UDF data expressing our preference for statistical
uncertainty over systematic uncertainty. 

\begin{table*}
\begin{tabular}{c|ccc}
                & $w_{1''}$ & $\theta_0$ &        $\Gamma$ \\
\hline
$25<V<26$        & $0.946 \pm 0.034 \pm 0.094 $   & $0.979 \pm 0.084 \pm 0.038$   & $
2.58 \pm
0.28$ \\
$26<V<27$        & $0.520 \pm 0.014 \pm 0.051 $   & $0.763 \pm 0.030 \pm 0.031$   & $2.43
\pm
0.12$ \\
$27<V<28$        & $0.296 \pm 0.053 \pm 0.028 $   & $0.60  \pm 0.10  \pm 0.024$   & $2.41
\pm
0.63$ \\
$28<V<29$       & $0.087 \pm 0.024 \pm 0.009$    & $0.438 \pm 0.042 \pm 0.016$   & $2.96
\pm 0.50$ \\
\hline
\end{tabular}
\caption{Best global estimates of $w_{1''}$, $\theta_0$ and $\Gamma$.}\label{tab:bestw}
\end{table*}


These results are consistent within computed error of a $\Gamma = 2.5$ in all cases.
Assuming this value gives us slightly different values of $\theta_0$ in Table
\ref{tab:finalw}.
\begin{table}
\begin{tabular}{c|cc}
                & $w(1'')$ & $\theta_0$ \\
\hline
$25<V<26$        & $0.889 \pm 0.041 \pm 0.088 $  & $0.955 \pm 0.077 \pm 0.038 $   \\
$26<V<27$        & $0.511 \pm 0.016 \pm 0.052 $  & $0.758 \pm 0.027 \pm 0.031 $   \\
$27<V<28$        & $0.298 \pm 0.076 \pm 0.030 $  & $0.598 \pm 0.091 \pm 0.024 $   \\
$28<V<29$        & $0.113 \pm 0.052 \pm 0.009 $  & $0.479 \pm 0.035 \pm 0.016 $   \\
\hline
\end{tabular}
\caption{Best global estimates of $w(1'')$ and $\theta_0$ assuming $\Gamma =
2.5$.}\label{tab:finalw}
\end{table}
The correlation length as a function of limiting magnitude \textit{V} is well fit by:
\begin{equation}
\theta_0(V) = 10^{-0.1(V-25.8)} \rm{arcsec};\ 26 < V < 29
\end{equation}
This is a factor of 40 larger than what we would expect from the extrapolation of a
purely gravitational
correlation function in equation \ref{eq:theta0}.

\begin{figure*}
\begin{minipage}[t]{0.49\linewidth}
\centering\includegraphics[width=\linewidth]{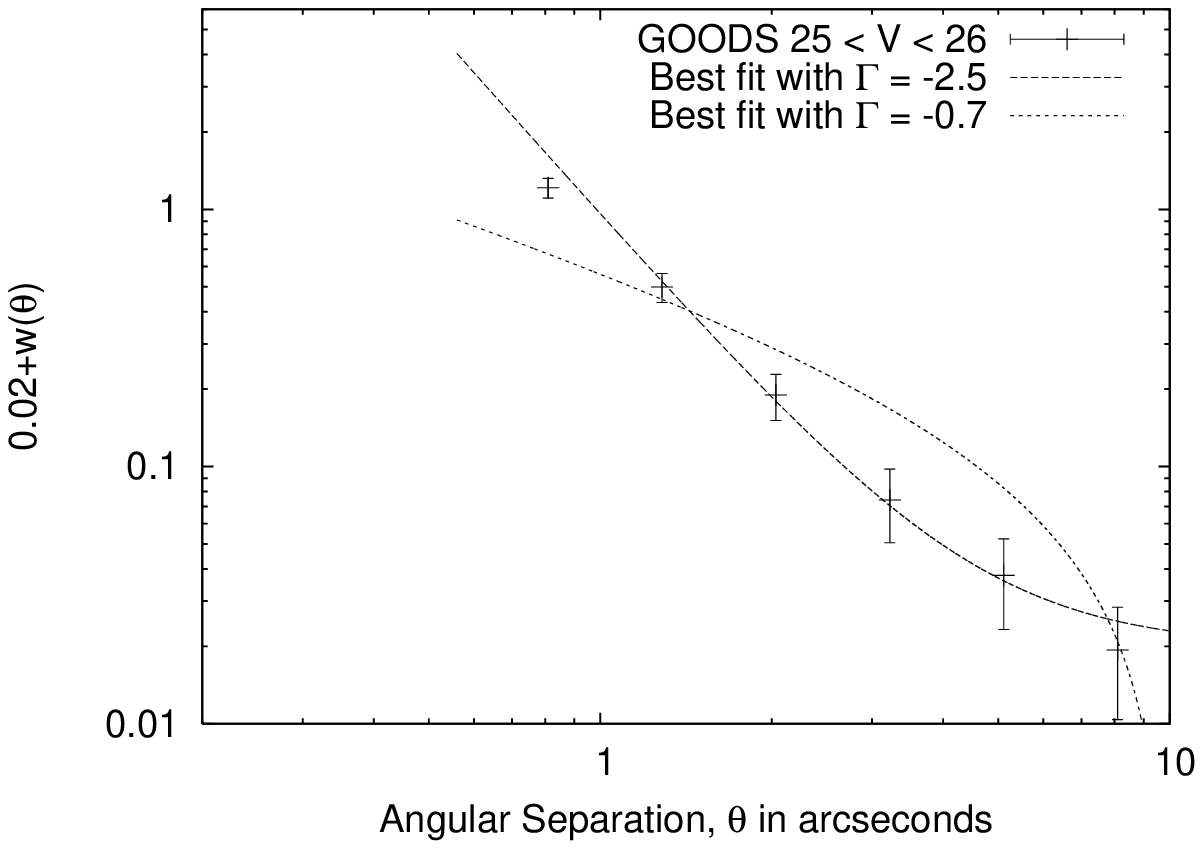}
\end{minipage} \hfill
\begin{minipage}[t]{0.49\linewidth}
\centering\includegraphics[width=\linewidth]{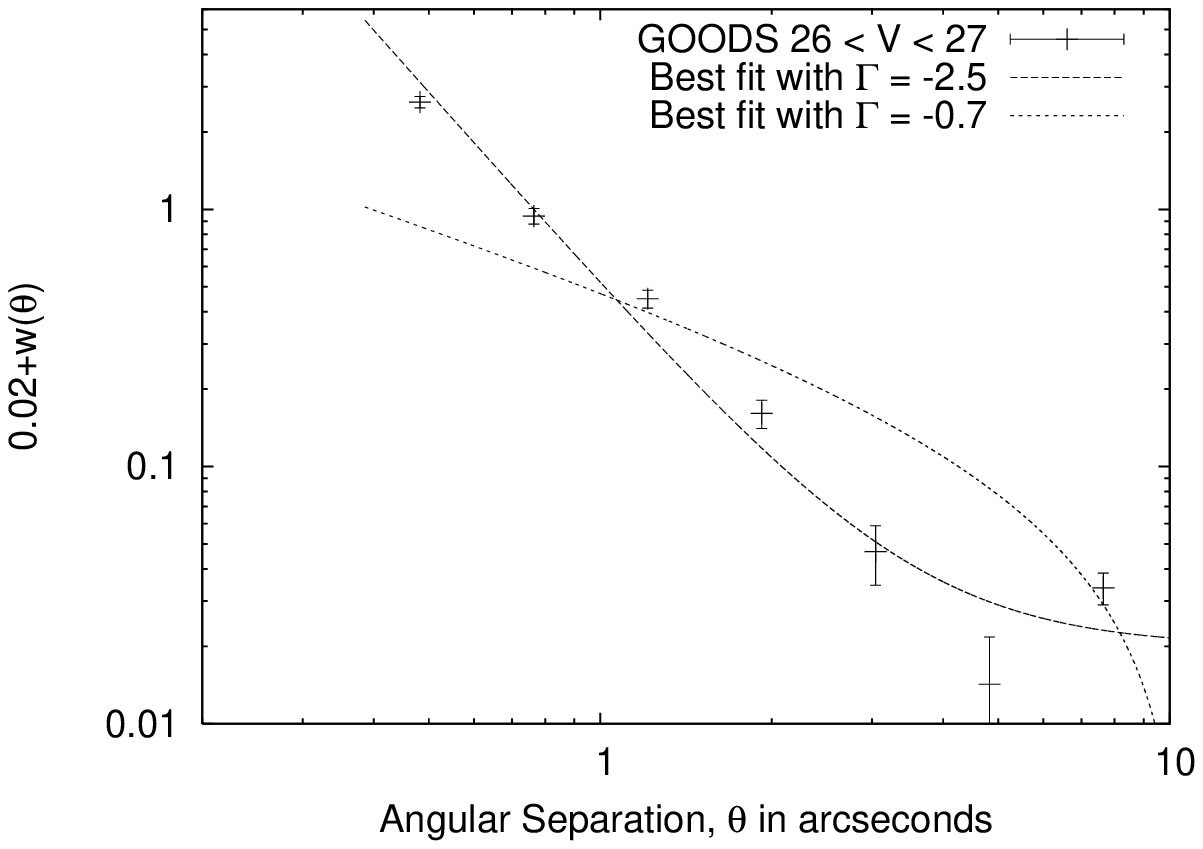}
\end{minipage} \hfill
\begin{minipage}[t]{0.49\linewidth}
\centering\includegraphics[width=\linewidth]{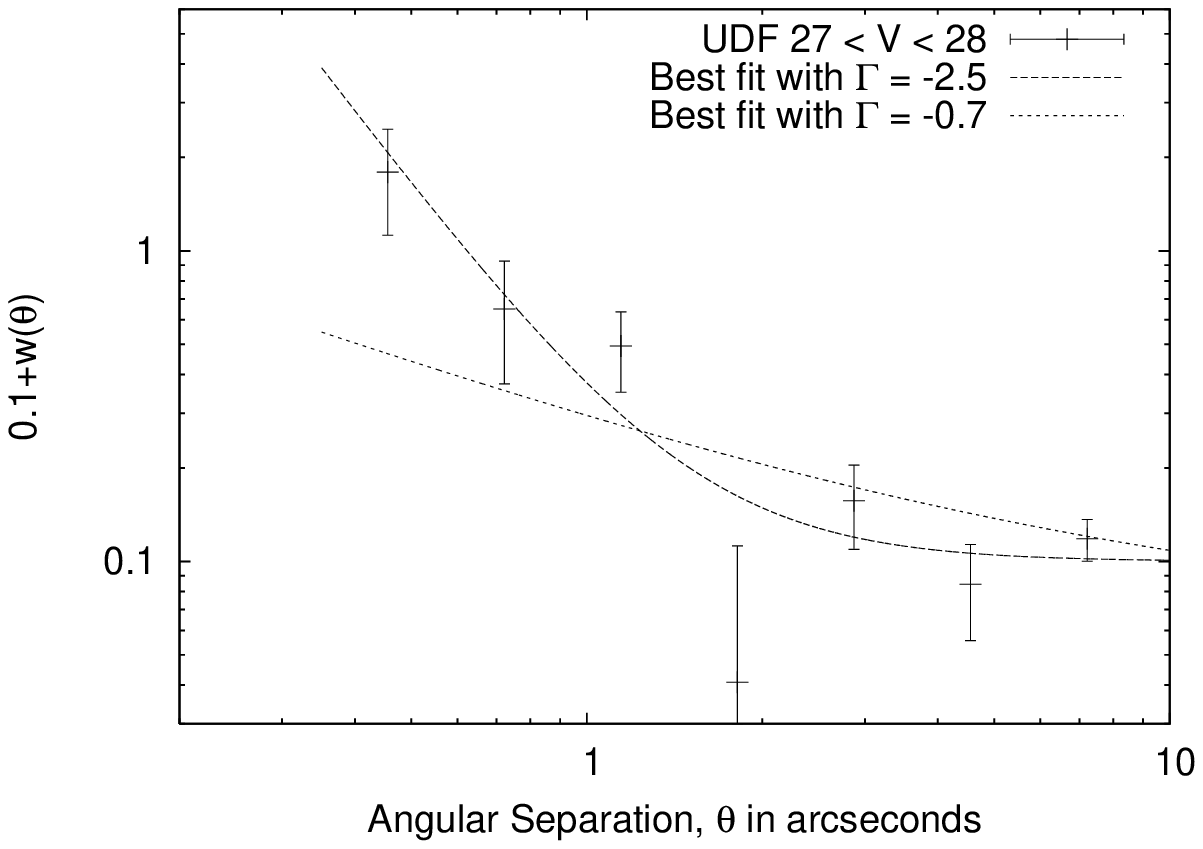}
\end{minipage} \hfill
\begin{minipage}[t]{0.49\linewidth}
\centering\includegraphics[width=\linewidth]{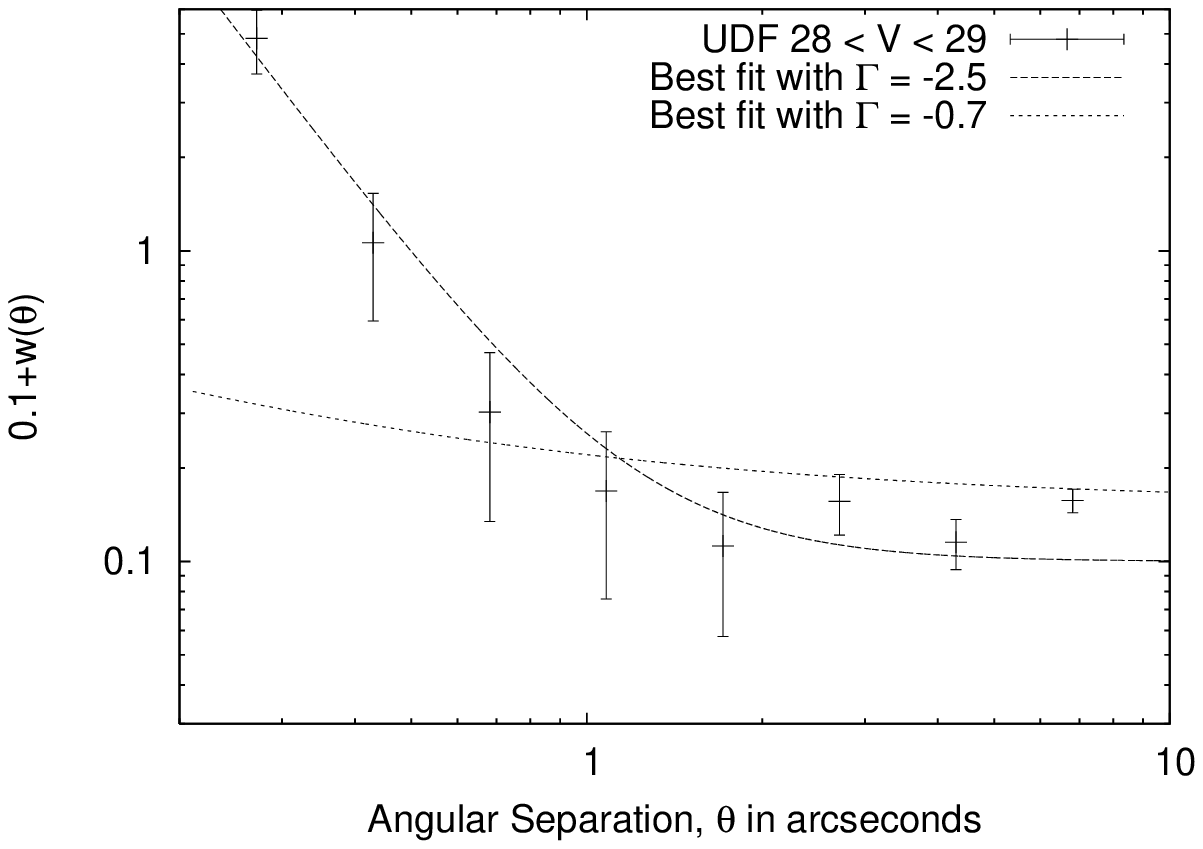}
\end{minipage} \hfill
\caption{The fitted correlation function in GOODS $25 < V < 26$ (top, left), GOODS $26 < V < 27$ (top, right), and UDF $27 < V < 28$ (bottom, left) and UDF $28 < V < 29$ (bottom, right). We show our best $\Gamma = 2.5$ and $\Gamma = 0.7$ fits. $w(\theta)$ is offset from zero by 0.02 (0.1) for GOODS (UDF) so that slightly negative points can be plotted on a log-log plot. This causes distortion in the plots at large angles. The $\Gamma = 0.7$ fits are actually $(\theta/\theta_0)^{-0.7}+\delta$ fits.}\label{fig:w}
\end{figure*}

We plot our results in Fig. \ref{fig:w} with the best $\Gamma = 2.5$ fit. We also plot the $\Gamma = 0.7$ fit to show that extending the SDSS power law to small scales vastly underestimates the number of close pairs. Note that we are plotting $0.02+w(\theta)$ for GOODS and $0.1+w(\theta)$ for UDF, because it allows for slightly negative points to be included in a log-log plot. Also note that the $\Gamma = 0.7$ fits are of $(\theta/\theta_0)^{-0.7}+\delta$, where $\delta$ is an independently fit parameter. $\delta$ is necessary, because different fit $\Gamma$'s lead to different $\sigma_0$ values and effectively offset $w(\theta)$.

The $\sigma_0$ we use slightly undervalues the true density, because we require our fitted $w(\theta)$ to always be positive. In a sense, our $\sigma_0$ represents the density of sources if there were no 'extra pairs' due to the correlation function. Given this expected discrepancy, our fitted results are consistent with being slightly less than $n_{sources} / Area$ and we do not print them here.

\subsection{Systematic Error}\label{subsec:sys}

Our methods produce systematic errors related to how our simulations differ from true
images. We trace these error to two effects: unrepresentative source profiles and
inadequate noise models. The effects manifest themselves as errors in $w_M$ and $w_O$. If our 
simulations are accurate, however, the effects should cancel out in our final 
measurement of $w_P$. But any discrepancies between our simulations and real images will prevent this cancellation and produce systematic errors in our measured $w_P$. We set generous upper limits on discrepancies in our estimate of systematic error and find that statistical error is still our major source of noise in $\theta_0$.

Object Extractor, like Source Extractor, subtracts a background profile from each source
so that the faint (below detection threshold) wings of each source do not make the
surrounding sources appear brighter. But for both the real and simulated images, this
process is imperfect and background subtraction may cause any source in the region
surrounding a given source to be recorded as brighter or dimmer than it truly is. 

Source profiles are generally larger than PSFs, and we were able to
reduce our error in PSF width to at most 2\%, so we focus on the source profiles. In
subsection \ref{subsec:profile}, we noted that the intensity of a typical source is at
most $0.05$ DETECT\_THRESH at the distances at which we search for pairs. This sets an
upper limit in the difference between observed and actual luminosity of roughly  $|\Delta
L / L| = 0.05$.

Source concentration is proportional to:
\begin{equation}
\sigma(L) \propto e^{\eta V(L)}
\end{equation}
and an uncertainty in $L$ will lead to a localized uncertainty in $\sigma$. This
localized uncertainty in $\sigma$ will produce or suppress pairs, directly altering
$w(\theta)$ at the scale of the improper background subtraction.

Applying $0.05$ fractional uncertainty in luminosity at the $\theta_0$ scale leads to a
background subtraction uncertainty, $\sigma_{bs}$, in $\theta_0$ of roughly:
\begin{eqnarray}
\sigma_{bs} &=& 0.05\ \left(1+w(\theta_0) \frac{\partial \sigma(L)}{\sigma(L)\ \partial L}
\frac{\partial \theta_0}{\partial w}(\theta = \theta_0)\right)\nonumber\\
&=& \frac{0.05 \times 2 \times \eta}{\log(10^{0.4})\Gamma}\theta_0
\end{eqnarray}

The effect of false detections is twofold. If the detections were randomly scattered
throughout the image, the extra $0.5\%$ sources would reduce the correlation function on
all scales by $0.5\%$. Any clustering of these sources could contribute to a positive
correlation function. But we do not observe strong clustering of false detections near
sources in subsection \ref{subsec:false} on the scales we probe here. We estimate that
false detections are at most twice as likely in the area $\theta < 2\theta_0$. This
implies that only about $0.1\%$ of sources would have an extra partner on these scales.
In subsection \ref{subsec:mult}, we find that roughly $10\%$ of sources have a partner on
these scales, so false detections are only a percent level source of error. We neglect
their contribution in our systematic error estimate.

A separate source of significant error derives from the use of the incompleteness factor, 
$\kappa$. We use this factor in our GOODS $27 < V <28$ and UDF $28 < V < 29$ measurements 
to counter the effects of incompleteness. The values we use reproduce the input correlation 
function in our simulation measurements, but we do not understand exactly how incompleteness
affects $w(\theta)$. The farther $\kappa$ is from its ideal value of unity, the more
uncertainty our use of $\kappa$ implies. We assign a fractional uncertainty in $\kappa$
equal to $0.5 (1 - \kappa)$ to produces generous error bars in our measurements of
$\theta_0$ in incomplete samples. This leads to an uncertainty $\sigma_{\kappa}$ in
$\theta_0$ of:
\begin{equation}
\sigma_{\kappa} = \frac{0.5 (1-\kappa) }{ \kappa \Gamma} \theta_0
\end{equation}

\begin{table}
\begin{tabular}{c|ccc}
Sample  &  $\sigma_{bs}$ & $\sigma_{\kappa}$ & $\sigma_{\theta_0}$ \\
\hline
$25 < V < 26$       & 0.038 & 0     & 0.038 \\
$26 < V < 27$       & 0.031 & 0     & 0.031 \\
$27 < V < 28$ GOODS & 0.022 & 0.070 & 0.074 \\
$27 < V < 28$ UDF   & 0.024 & 0     & 0.024 \\
$28 < V < 29$       & 0.011 & 0.011 & 0.016 \\
\hline
\end{tabular}
\caption{Systematic errors due to background subtraction,  }\label{tab:syserr}
\end{table}

We do not apply systematic error to $\Gamma$, because $\Gamma$ is highly dependent on the
small number of close pairs, and random errors dominate systematic errors.

In this paper, we do not address the issue of cosmic variance. While each field is many times larger than our critical distance of roughly one arcsecond, cosmic source densities can vary on large scales and the effect of such variation on small scale clustering is unclear. It is likely that different clusters of sources are separated by cosmological distances along the line of sight. So by taking an angular measurement, we may be averaging out cosmic variance. There is no statistically significant variation in $\theta_0$, $\Gamma$ or source density $\sigma_0$ between GOODS North and GOODS South, and we have no reason to believe that cosmic variance is a significant effect.

%
%

\subsection{Multiplicity Fractions}\label{subsec:mult}
Measuring the fraction of sources in close pairs is another way to study clustering. It
allows us to compare the correlation length with the average separation of sources. These
fractions also give us detailed estimates of how many pairs we have in the sky.

\begin{figure}
\centering\includegraphics[width=\linewidth]{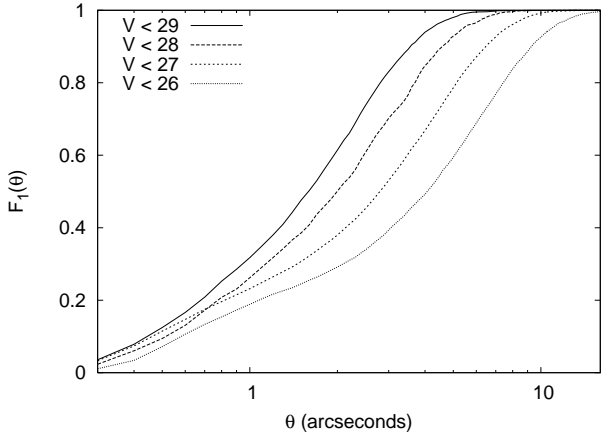}
\caption{$F_1(\theta)$, the fraction of sources with a partner within $\theta$ (left).
The $V < 27$ sources are taken from GOODS South and the $V > 27$ sources are taken from
UDF.}\label{fig:fpairs}
\end{figure}

In Fig. \ref{fig:fpairs} we see $F_1(\theta)$, the fraction of sources with one neighbor
within $\theta$ of them. In an unclustered sample, we would have:
\begin{equation}
F_{1 u}(\theta) = 1 - e^{-\pi \sigma_0 (\theta^2 - \theta^2_{min})}
\end{equation}
While in a clustered sample, we have:
\begin{equation}
F_{1 c}(\theta) = 1 - e^{-\pi \sigma_0 ((\theta^2 - \theta^2_{min}) + \theta_0^\Gamma
/(2-\Gamma)(\theta^{2-\Gamma}-\theta_{min}^{2-\Gamma} ))}
\end{equation}

In our samples, the average separation is $3'' < (\pi \sigma_0)^{-1/2} < 8''$. On scales
smaller than this, $F_1$ would go as $\theta^2$ if the samples were unclustered. Instead,
in Fig. \ref{fig:fpairs} we see a steep initial rise with many close pairs and then a
flattening out at $\theta_0$ as $w(\theta)$ ceases to dominate.

Finally, the $F_1$ function allows us to estimate the number of pairs that we see on the
sky. In Table \ref{tab:fpairs}, we see a roughly exponentially increasing number of pairs
within $2 \theta_0$ as we go to fainter magnitudes which cuts off at $V \approx 28$. The
failure to find more high magnitude pairs could be due to the fact that the $V > 28$
pairs would be on scales very near the resolution limit of the instrument. In any event,
if we are to use the FSCF with roughly HST-like space telescopes, we must probe faint sources $V > 25$ to get good statistics.

\begin{table}
\begin{tabular}{c|ccc}
Sample  &  $\sigma_0$ degree$^{-2}$ & $\theta_0$ & $N_{pairs}(2\theta_0)$ degree$^{-2}$ \\
\hline
$25 < V < 26$ &  $1.4 \times 10^5$ & $1.06''$  & $1.6 \times 10^4$ \\
$26 < V < 27$ &  $2.8 \times 10^5$ & $0.880''$  & $4.3 \times 10^4$ \\
$27 < V < 28$ &  $5.0 \times 10^5$ & $0.64''$  & $1.1 \times 10^5$ \\
$28 < V < 29$ &  $7.8 \times 10^5$ & $0.493''$  & $1.3 \times 10^5$ \\
\hline
\end{tabular}
\caption{Integrated source and pair counts (per square degree) in our catalogs. The  $V <
27$ sources are taken from GOODS South and the $V > 27$ sources are taken from UDF.
}\label{tab:fpairs}
\end{table}
The integral counts are roughly consistent with:
\begin{equation}
N(V) = 10^{0.4(V-12.7)-0.03(V-25.3) /\rm{deg}^2}
\end{equation}
This formula includes incompleteness in our datasets.

\section{Discussion}

In this paper we have presented measurements of the two point angular correlation
function for faint ($25 < V < 29$) sources. This measurement has been validated by
extensive numerical simulation. The observed correlation function is consistent with:

\begin{equation}\label{eq:w_est}
w(\theta) = \left(\frac{\theta}{\theta_0(V)}\right)^{-2.5};\ \theta_0(V) =
10^{-0.1(V-25.8)} \rm{arcsec}
\end{equation}

This measurement shows that the FSCF has a much steeper slope and larger normalization than the SDSS angular 
correlation function for LRGs would suggest if extrapolated. This is not surprising 
since we are looking at smaller scale physics, at bluer sources and likely at different 
redshifts. This measurement is not a direct extension of the LRG work, but instead
an analogous measurement for a different dataset.

There are many possible uses for this measurement. First, our observation of this galactic scale 
correlation function can be compared with numerical simulations of galaxy formation
that include gravitational clustering, gas dynamics, star formation, etc. In this
application it is important that the simulation data be processed in the same way as the
observation. Alternately, it is possible to reanalyze this data with a method that
derives from the simulation.

Our measurement errors are limited to roughly $10\%$ by the data. They could be reduced
substantially with larger samples as has happened with SDSS and the bright source
$w(\theta)$. It is unlikely that
significantly more deep field type data can be mined from the existing HST instruments.
However the Wide Field Camera 3 should be deployed this year and will provide 7 square
arcminute exposures of high resolution data with similar limiting magnitudes to the
Advanced Camera for Surveys \citep{WFC32006}. Looking ahead, the James Webb Space
Telescope (JWST) is scheduled to launch in 2013 and will provide 5 square arcminute
frames with $0.1''$ resolution and a limiting magnitude of at least $K \approx 25$ ($V
\approx 30$ for galactic sources) \citep{JWST2006}. JWST should find many high redshift galaxies. In Table \ref{tab:surveys}, we see
that if we could study the FSCF across many fields to obtain several square degrees of
data, we could vastly improve our measurements of $w(\theta)$

Table \ref{tab:surveys} also shows that several ground-based projects will provide enough
data to overcome the small amplitude of the FSCF on arcsecond scales and yield significant
measurements. The Dark Energy Survey \citep{DES2005} will produce a precise measurement
for the $V < 24$ sources that we do not study here. The Panoramic Survey Telescope and
Rapid Response System (Pan-STARRS) \citep{PANS2007} and the Large Synoptic Survey
Telescope (LSST)\citep{LSST2008} could probe 30,000 $deg^2$ and 20,000 $deg^2$
respectively and measure the FSCF of the brighter sources than those we study here with
precision of
$\approx 10^{-4}$.

But the superior resolution and huge area of possible future space-based surveys makes
them the ideal candidates for this method. The Supernovae Acceleration Probe (SNAP) would
measure roughly 1,000 $deg^2$ with a limiting magnitude of approximately 28
\citep{SNAP2005}. Destiny has a similar project goal \citep{BENF2006}. These enormous
surveys would yield FSCF results that directly study $V < 28$ sources on the pertinent
subarcsecond scales, reducing the statistical noise we encounter here by a factor of
several hundred.

\begin{table*}
\begin{tabular}{ccccccc}
Dataset           & Area ($deg^2$)   & Limiting \textit{V} Magnitude & $\theta_{min}$ &
$N_{sources}$
 & $w_{faint}(\theta_{min})$ & $\sigma_{w_{faint}})/w_{faint}$ \\
\hline
HST-WFC3   & $5$              & $27$                 & $0.5''$        & $3 \times 10^6$
 & $2.8$                    & $3 \times 10^{-3}$ \\
JWST       & $5$              & $30$                 & $0.5''$        & $4 \times 10^7$
 & $0.50$                   & $9 \times 10^{-4}$ \\
DES        & $5000$           & $24$                 & $1.5''$        & $2 \times 10^8$
 & $1.0$                    & $1 \times 10^{-3}$ \\
Pan-STARRS & $30000$          & $26.5$               & $1.5''$        & $1 \times
10^{10}$ & $0.24$                   & $2 \times 10^{-4}$ \\
LSST       & $30000$          & $27.5$               & $1.5''$        & $2 \times
10^{10}$ & $0.14$                   & $1 \times 10^{-4}$ \\
SNAP       & $1000$           & $28 $                & $0.4''$        & $1 \times 10^9$
 & $2.8$                    & $2 \times 10^{-4}$ \\
\hline
\end{tabular}
\caption{Characteristics of upcoming datasets and their potential to measure $w(\theta)$.
The area and limiting magnitude of the HST-WFC3 and JWST  are estimated, and these
datasets will not be contiguous or uniform surveys. For JWST, we estimate the limiting V
magnitude given a $K \approx 25$ limit and galactic sources. $\theta_{min}$ is an
estimated minimum $\theta$ at which we could reasonably make $w(\theta)$ measurements and
is equal to roughly three times the PSF width. $w_{faint}(\theta_{min})$ is the size of
the correlation function in the faintest single magnitude bin assuming Eq. \ref{eq:w_est}
holds. $\sigma_{w_{faint}}/w_{faint}$ is the fractional statistical uncertainty in this
measurement assuming we bin pairs with $\theta_{min} < \theta < 1.58\ \theta_{min}$.
}\label{tab:surveys}
\end{table*}

To learn more about these faint sources, we must know their distance. The sources we
discuss in this paper are roughly 5 magnitudes too dim for spectroscopy, 3 magnitudes too
dim for traditional weak lensing measurements and roughly 2 magnitudes beyond the range
where one can rely on training sets to produce accurate photometric redshifts.

Fortunately, lensing enables a new approach to measuring the distance to these sources.
Consider first, a field of sources that is gravitationally lensed by large scale
structure. The density of a given population of sources will vary inversely with the
magnification, $\mu$. But amplification also brings faint sources above the detection
threshold, and if the number of sources near the detection threshold goes as
$L^{-\beta}$, then the density of total sources goes as $\mu^{\beta-1}$. In practice
$\beta\approx 0.8$ and the effects nearly cancel. This and the fact that amplification on
cosmic scales is only a few percent make this measurement exceedingly difficult.

But measuring the effect of shear on the FSCF is relatively straight forward and
achievable using the large datasets mentioned above. A uniform shear, $\gamma$ breaks the
azimuthal symmetry of the FSCF and, for a power law, the FSCF becomes:
\begin{equation}
w(\theta, \phi) = w(\theta)(1+\gamma \Gamma \cos(2\phi))
\end{equation}
where $w(\theta)$ is the unsheared FSCF, and $\phi$ is the angle between the
source-source separation vector and the axis of shear.

If we have high resolution data with a limiting magnitude of roughly $28$, we expect
$10^5$ close pairs per square degree. With a 1,000 square degree field, we would be
statistically limited to measuring to measuring $w(\theta)$ at the $10^{-4}$ level of
precision. A $\gamma$ of $0.01$ could in turn be measured with roughly percent precision.
Instead of using this method to measure $\gamma$, however, we will use the superior
measurements of shear gained from traditional weak lensing of brighter sources with
calibrated photometric redshifts to determine the redshift distribution of the faintest
sources in the sky.

Cosmic shear is in many ways the most difficult type of lensing measurement to make, and
it is likely that the `Pair Lensing' we describe here will follow a similar
observational path to traditional weak lensing, first being observed around clusters and
galaxies and then being observed in large field.

We plan to explore the FSCF in more detail in future papers. In paper II of this series
we will compute the three point correlation function for these faint sources. In paper
III we will discuss the theory of gravitational lensing on the FSCF in more details and
apply our results to the three environments mentioned above. In paper IV we will use
existing data to attempt to observe this lensing phenomenon and examine in depth the
possibility of making a more precise measurement in a larger dataset.

 \section{Acknowledgments}
We thank Phil Marshall and David Hogg for useful discussions and guidance. We thank the
GOODS team, particularly Leonidas Moustakas and Victoria Laidler, for consultation on
source extraction and software. This work was supported by the NSF under award
AST05-07732 and in part by the U.S Department of Energy under contract number
DE-AC02-76SF00515.
\bibliography{EFM}
\label{lastpage}

\end{document}